\pdfoutput=1
\RequirePackage{ifpdf}
\ifpdf 
\documentclass[pdftex]{sigma}
\else
\documentclass{sigma}
\fi

\usepackage{mathrsfs}
\usepackage{bm}
\usepackage{tikz}
\usepackage{cite}

\DeclareMathAlphabet\mathbfcal{OMS}{cmsy}{b}{n}

\def\be{\begin{equation}}
\def\ee{\end{equation}}

\def\ri{{\rm i}}

\def\XXint#1#2#3{{\setbox0=\hbox{$#1{#2#3}{\int}$}
 \vcenter{\hbox{$#2#3$}}\kern-.5\wd0}}

\newcommand{\p}{\partial}

\newcommand{\re}{\mbox{e}}

\newcommand{\Ecal}{{\cal E}}

\newcommand{\Hcal}{{\cal H}}

\newcommand{\leng}{{N}}

\numberwithin{equation}{section}

\begin{document}

\newcommand{\arXivNumber}{2010.10615}

\renewcommand{\thefootnote}{}

\renewcommand{\PaperNumber}{025}

\FirstPageHeading

\ShortArticleName{Some Algebraic Aspects of the Inhomogeneous Six-Vertex Model}

\ArticleName{Some Algebraic Aspects \\of the Inhomogeneous Six-Vertex Model\footnote{This paper is a~contribution to the Special Issue on Mathematics of Integrable Systems: Classical and Quantum in honor of Leon Takhtajan. The full collection is available at \href{https://www.emis.de/journals/SIGMA/Takhtajan.html}{https://www.emis.de/journals/SIGMA/Takhtajan.html}}}

\Author{Vladimir V. BAZHANOV~$^{\rm a}$, Gleb A.~KOTOUSOV~$^{\rm b}$, Sergii M.~KOVAL~$^{\rm a}$\\ and Sergei L.~LUKYANOV~$^{\rm cd}$}

\AuthorNameForHeading{V.V.~Bazhanov, G.A.~Kotousov, S.M.~Koval and S.L.~Lukyanov}

\Address{$^{\rm a)}$~Department of Theoretical Physics, Research School of Physics,\\
\hphantom{$^{\rm a)}$}~Australian National University, Canberra, ACT 2601, Australia}
 \EmailD{\href{mailto:vladimir.bazhanov@anu.edu.au}{vladimir.bazhanov@anu.edu.au}, \href{mailto:sergii.koval@anu.edu.au}{sergii.koval@anu.edu.au}}

\Address{$^{\rm b)}$~DESY, Theory Group, Notkestrasse 85, Hamburg, 22607, Germany}
\EmailD{\href{mailto:gleb.kotousov@desy.de}{gleb.kotousov@desy.de}}

\Address{$^{\rm c)}$~NHETC, Department of Physics and Astronomy, Rutgers University,\\
\hphantom{$^{\rm c)}$}~Piscataway, NJ 08855-0849, USA}
\EmailD{\href{mailto:sergei@physics.rutgers.edu}{sergei@physics.rutgers.edu}}

\Address{$^{\rm d)}$~Kharkevich Institute for Information Transmission Problems, Moscow, 127994, Russia}

\ArticleDates{Received October 30, 2020, in final form February 26, 2021; Published online March 16, 2021}

\Abstract{The inhomogeneous six-vertex model is a 2$D$ multiparametric integrable statistical system. In the scaling limit it is expected to cover different classes of critical behaviour which, for the most part, have remained unexplored. For general values of the parameters and twisted boundary conditions the model possesses ${\rm U}(1)$ invariance. In this paper we discuss the restrictions imposed on the parameters for which additional global symmetries arise that are consistent with the integrable structure. These include the lattice counterparts of~${\mathcal C}$, ${\mathcal P}$ and ${\mathcal T}$ as well as translational invariance. The special properties of the lattice system that possesses an additional ${\mathcal Z}_r$ invariance are considered. We also describe the Hermitian structures, which are consistent with the integrable one. The analysis lays the groundwork for studying the scaling limit of the inhomogeneous six-vertex model.}

\Keywords{solvable lattice models; Bethe ansatz; Yang--Baxter equation; discrete symmet\-ries; Hermitian structures}

\Classification{16T25; 52C26; 81T40; 82B20; 82B23}

\begin{flushright}
\begin{minipage}{85mm}
\it Dedicated to Professor Leon Armenovich Takhtajan\\ on the occasion of his 70$^{th}$ birthday
\end{minipage}
\end{flushright}

\renewcommand{\thefootnote}{\arabic{footnote}}
\setcounter{footnote}{0}

\section{Introduction}

The six-vertex model has played a
central r\^{o}le in the theory of exactly solvable
models in~sta\-tistical mechanics.
Its origins go back to the Pauling ice model~\cite{Pauling}
and it was later employed in~the description of
a two-dimensional ferroelectric
(for a review see~\cite{Lieb:1980ix,Reshetikhin:2010}).
The free energy per site in the
thermodynamic limit for a special ``square ice'' case
was calculated by Lieb~\cite{Lieb:1967zz}.
The result was then extended to the more general ferroelectric model
by Sutherland, Yang and Yang
\cite{Sutherland:1967a,CPYang:1967,
Sutherland:1967zz}.
All these works dealt with the homogeneous statistical system defined on the square lattice.
They used the technique pioneered by Bethe
in~\cite{Bethe:1931}, which allows one to compute the eigenvectors and eigenvalues of the transfer matrix.
Later on, Baxter in the work~\cite{Baxter:1971cs} considered the case where the local Boltzmann weights
vary from site to site.
He found the most general inhomogeneous six-vertex model, which is still solvable
by means of the
Bethe ansatz technique.
This was an important step
that led to the concept of commuting transfer matrices.
Subsequently, in his
seminal paper on the exact solution of the
eight-vertex model~\cite{Baxter:1972hz}, Baxter developed
the method of commuting transfer matrices, laying
the foundation of the modern theory of integrable quantum systems.
The approach was further extended
by Faddeev, Sklyanin and Takhtajan
within the quantum inverse scattering method (QISM)
\cite{Faddeev:1979gh,Takhtajan:1979iv}.

\begin{figure}[t]
\centering
\scalebox{0.85}{
\begin{tikzpicture}
\draw[thick] (0,-1.5) -- (6.5,-1.5);
\draw[thick] (7.5,-1.5) -- (13,-1.5);
\draw[thick] (2,-0.3) -- (2,-2.7);
\draw[thick] (4,-0.3) -- (4,-2.7);
\draw[thick] (6,-0.3) -- (6,-2.7);
\draw[thick] (8,-0.3) -- (8,-2.7);
\draw[thick] (10,-0.3) -- (10,-2.7);
\draw[thick] (12,-0.3) -- (12,-2.7);
\draw[fill = black] (1,-1.5) circle (.6ex);
\draw [thick] ([shift=(-180:0.4cm)]2,-1.5) arc (-180:-90:0.4cm);
\draw [thick] ([shift=(-180:0.4cm)]4,-1.5) arc (-180:-90:0.4cm);
\draw [thick] ([shift=(-180:0.4cm)]6,-1.5) arc (-180:-90:0.4cm);
\draw [thick] ([shift=(-180:0.4cm)]8,-1.5) arc (-180:-90:0.4cm);
\draw [thick] ([shift=(-180:0.4cm)]10,-1.5) arc (-180:-90:0.4cm);
\draw [thick] ([shift=(-180:0.4cm)]12,-1.5) arc (-180:-90:0.4cm);
\node at (-0.4,-1.5) {$\zeta$};
\node at (2.35,-2.5) {$a_N$};
\node at (2.1,-3.15) {$\eta_N$};
\node at (4.55,-2.5) {$a_{N-1}$};
\node at (4.3,-3.15) {$\eta_{N-1}$};
\node at (10.30,-2.5) {$a_{2}$};
\node at (10.1,-3.15) {$\eta_{2}$};
\node at (12.30,-2.5) {$a_1$};
\node at (12.1,-3.15) {$ \eta_1$};
\node at (2.35,-0.4) {$b_N$};
\node at (4.55,-0.4) {$b_{N-1}$};
\node at (10.30,-0.4) {$b_{2}$};
\node at (12.30,-0.4) {$b_1$};
\node at (3,-1.2) {$ c_N$};
\node at (5,-1.2) {$c_{N-1}$};
\node at (9,-1.2) {$c_{3}$};
\node at (11,-1.2) {$c_2$};
\node at (12.7,-1.2) {$c_1$};
\node at (0.5,-1.2) {$c_1$};
\node at (1.5,-1.2) {$c_1$};
\node at (7,-0.5) {$\cdots$};
\node at (7,-1.5) {$\cdots$};
\node at (7,-2.5) {$\cdots$};
\end{tikzpicture}
}
\caption{A graphical representation of the transfer matrix
$(\mathbb{T}(\zeta))^{b_N b_{N-1}\dots b_1}_{a_N a_{N-1}\dots a_1}$ defined in~\eqref{tmat}.
The black dot corresponds to the boundary twist $\omega^{c_1}$.
Summation over the spin indices assigned to internal edges is assumed.
\label{pic2}}
\end{figure}
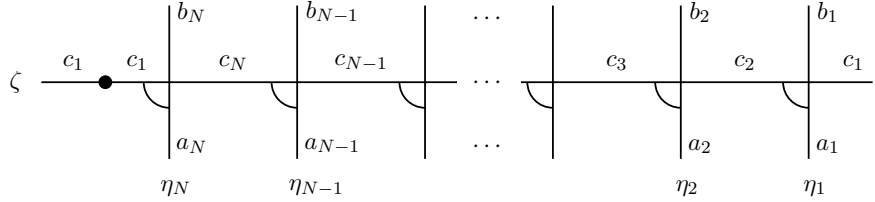

The local Boltzmann weights in the six-vertex model
are expressed through the
elements of the $R$-matrix, which is the
trigonometric solution of
the Yang--Baxter equation. The latter depends essentially
on two parameters~-- the anisotropy and the spectral parameter.
The inhomogeneous model considered in~\cite{Baxter:1971cs,Baxter:1972hz}
is defined on the square lattice, where each
horizontal and vertical line is associated with its own
variable. The local Boltzmann weights at each vertex
have the same value of the anisotropy parameter, while the spectral
parameter coincides with the ratio of the two
 variables assigned to the lines
passing through the vertex.
In this setting the row-to-row transfer matrix
${\mathbb T}$
for the lattice with $N$ vertical columns
will depend on the
variable~$\zeta$, associated with the horizontal line, as well as
the set $\{\eta_1,\eta_{2},\dots,\eta_{N}\}$,
corresponding to the vertical lines (see Fig.~\ref{pic2}).
The Yang--Baxter equation implies that for the (quasi-)periodic
boundary conditions the transfer
matrices with different values of $\zeta$, but with the same sets $\{\eta_J\}$
form a commuting family
\begin{gather*}
\big[{\mathbb T}(\zeta),
{\mathbb T}(\zeta')\big]=0.
\end{gather*}
The inhomogeneous six-vertex model turns out to be a multiparametric, integrable, statistical system
 that depends on the variables associated with the vertical lines, $\eta_J$,
those labeling the horizontal lines and the anisotropy parameter.
The latter will be referred to as $q$.

The most studied case of the model is
when the set of local Boltzmann weights
is the same at each vertex.
Then
$\eta_1=\eta_2=\cdots=\eta_N$
and the transfer matrix commutes with the spin $\frac{1}{2}$ Heisenberg
$XXZ$ Hamiltonian. The anisotropy parameter
of the spin chain, which is usually denoted as $\Delta$, is related to $q$ as
$\Delta=\frac{1}{2}\big(q+q^{-1}\big)$.
It determines the most interesting
 physical properties of the model. In particular when $-1<\Delta<1$
or, equivalently, $q$ is unimodular ($|q|=1$)
the system turns out to be critical and its scaling behaviour is governed by
a conformal field theory (CFT) of
a massless Gaussian field~\cite{Luther:1975wr,Kadanoff:1978pv,Alcaraz:1987zr}.
Perhaps the simplest inhomogeneous six-vertex model
corresponds to a staggering of the
variables associated with the vertical lines of~the lattice
such that
$\eta_1=\eta_3=\cdots$, $\eta_2=\eta_4=\cdots$,
while $\eta_1\ne \eta_2$ and a similar staggering for the variables
assigned to the horizontal lines. In the case when
$\eta_1/\eta_2>0$ and $|q|=1$, with a~properly defined scaling limit
where $\eta_1/\eta_2\to 0^+$ (or $+\infty$) and $N\to\infty$,
the lattice model exhibits universal behaviour that is described by the massive Thirring/sine--Gordon model
\cite{Japaridze:1982yc,Destri:1987}.\footnote{Generating a correlation length in a critical homogeneous lattice system
via the introduction of an alternating set of inhomogeneities has been used in a variety of
contexts see, e.g.,~\cite{Polyakov:1983tt,Faddeev:1985qu}.}
The~situation is different if the ratio of $\eta_1$ to $\eta_2$ is negative.
For $\eta_1/\eta_2=-1$ and $|q|=1$ the system is critical and its important feature
is the presence of a ${\cal Z}_2$
symmetry. For this reason the model is sometimes referred to as the ${\cal Z}_2$ invariant inhomogeneous
six-vertex model.
Various types of universal behaviour occur depending on whether ${\rm arg}\big(q^2\big)\in\big(0,\frac{\pi}{2}\big)$
or~$\big(\frac{\pi}{2},\pi\big)$. The latter case was considered in~\cite{IJS2}.
The former was the subject matter of the works
\cite{Jacobsen:2005xz,Ikhlef:2008zz,Ikhlef:2011ay,Frahm:2012eb,Candu:2013fva,Frahm:2013cma,Bazhanov:2019xvy}.
It was originally observed in~\cite{Jacobsen:2005xz,Ikhlef:2008zz} that
there is a continuous component in the spectrum of~conformal dimensions for the model.
 In~\cite{Bazhanov:2019xvy} it was pointed out that
the algebra of extended conformal symmetry underlying the critical behaviour is the
$W_\infty$ algebra.

\looseness=1
The identification of the critical surfaces in the space of couplings of
the inhomogeneous six-vertex model
as well as a description of the corresponding
universality classes is, without doubt, an interesting problem.
However, apart from the few cases mentioned above not much work
has been done in this direction. An evident starting point to
approach the problem
would be to explore the restrictions on the parameters
which give rise to extra symmetries in the model.
The~latter would include
translational invariance as well as global involutions such as ${\cal C}$, ${\cal P}$ and~${\cal T}$,
which are common attributes of a local QFT.
There may also be other global symmetries, e.g., the~${\cal Z}_2$ invariance
of the staggered model with $\eta_J/\eta_1=(-1)^{J-1}$. Of~course, the
identification of~the universality classes requires more than just an analysis of the symmetries.
The integrability of the model makes possible a detailed quantitative study of the scaling limit.
The goal of this work is to describe the
integrable structure underlying the general inhomogeneous six-vertex model
as well as its interplay with the global sym\-metries.

The paper is organized as follows.
In Section~\ref{sec2} we set-up the notation and
give a short summary of the results of~\cite{Baxter:1971cs}.
We also present the construction of the transfer matrix as well as its eigenstates (Bethe states)
within the framework of the QISM.
The commuting family for the inhomogeneous six-vertex model,
apart from $\mathbb{T}(\zeta)$,
contains
other important members~-- the so-called
Baxter ${Q}$-operators~\cite{Baxter:1972hz}.
Their construction
along the lines of~\cite{Bazhanov:1996dr,Bazhanov:1998dq}
is presented in~Section~\ref{sec3}.

Section~\ref{sec4}
is devoted to a discussion of the ``charge'' conjugation ${\cal C}$,
parity inversion ${\cal P}$ and time reversal transformation ${\cal T}$.
We describe the conditions under which these conjugations
are consistent with the integrable structure in the sense that they
preserve the family of commuting operators.

For a generic set of the inhomogeneities
the transfer matrix and $Q$-operators are not Hermitian
w.r.t.\ the standard matrix Hermitian conjugation.
In Section~\ref{sec5}
a family of
Hermitian structures is introduced,
which are also consistent with the integrable
structure of the model.
The norm of the Bethe states associated with these Hermitian structures
can be computed via the formula
originally conjectured by Gaudin, McCoy and Wu~\cite{Gaudin:1981cyg}
for the homogeneous case
and then extended and proven, using the formalism of the QISM, for
the general inhomogeneous model by Korepin~\cite{Korepin:1982gg}.

\looseness=1
Further restrictions on the parameters of the model,
which lead to the presence of additional global symmetries,
are discussed in Sections~\ref{sec6} and \ref{sec7}. Among these is
the invariance w.r.t.~lattice translations.
The latter appears when the number of columns of the lattice is
divi\-sible by some integer $r\ge1$ and the inhomogeneities $\eta_J$
satisfy the $r$-site periodicity conditions $\eta_{J+r}=\eta_J$.
In~this case the commuting family of the model
contains $r$ Hamiltonians, which
are given by a~sum of
local operators.
By the latter we mean that they are built from the local spin operators acting
on $r+1$ consecutive sites of the lattice.
Section~\ref{sec7} discusses the lattice system with the
$\{\eta_J\}$ being further restricted such that
$\eta_J/\eta_1=\re^{2\pi\ri(J-1)/r}$. Then the model possesses an extra~${\cal Z}_r$ cyclic group symmetry.
The last section contains a collection of~explicit
formulae for the model with $r=1$ (homogeneous) and $r=2$ (${\cal Z}_2$ invariant).
They lay the groundwork for a~detailed analysis of the scaling limit for these two cases,
which is perfor\-med~in~\cite{Bazhanov:2020}.

\section{The inhomogeneous six-vertex model\label{sec2}}
Consider the square lattice
with $\leng$ vertical columns.
Each edge of the lattice can be in one of two states,
labeled as $\pm1$.
Introduce the $R$-matrix of the six-vertex model, whose non-zero entries
are given by
\begin{gather}
R^{++}_{++}(\zeta)=R^{--}_{--}(\zeta)=q+q^{-1}\zeta,\qquad
R^{-+}_{+-}(\zeta)=-\big(q-q^{-1}\big)\zeta,\nonumber\\
R^{+-}_{+-}(\zeta)=R^{-+}_{-+}(\zeta)=1+\zeta,\qquad
R^{+-}_{-+}(\zeta)=q-q^{-1},\label{rmat}
\end{gather}
where $\zeta$ and $q$ are arbitrary parameters.
Let $a_1,a_2,\dots,a_\leng$ denote the states of a row of
$\leng$ vertical edges of the lattice and $b_1,b_2,\dots,b_\leng$
be the states on the row above. The
elements of~the row-to-row transfer matrix of the integrable inhomogeneous
six-vertex model is defi\-ned~as
\begin{gather}
({\mathbb T}(\zeta)\big)_{a_{N} a_{N-1}\dots a_1}^{b_{N}b_{N-1}\dots b_1}
= q^{-\frac{N}{2}}\sum_{c_1c_2\cdots c_{N}=\pm}\omega^{c_1}
R\big(q\zeta/\eta_N\big)_{c_{1}a_{N}}^{c_{N}b_{N}}
R\big(q\zeta/\eta_{N-1}\big)_{c_{N}a_{N-1}}^{c_{N-1} b_{N-1}}\cdots\nonumber
 \\ \hphantom{({\mathbb T}(\zeta)\big)_{a_{N} a_{N-1}\cdots a_1}^{b_{N}b_{N-1}\cdots b_1}
= q^{-\frac{N}{2}}\sum_{c_1c_2\cdots c_{N}=\pm}}
\cdots R\big(q\zeta/\eta_{2}\big)_{c_{3} a_{2}}^{c_{2} b_{2}}R
\big(q\zeta/\eta_1\big)_{c_{2}a_{1}}^{c_{1} b_{1}}.
\label{tmat}
\end{gather}
Here we have imposed twisted boundary conditions depending on the
parameter $\omega$, while
$\eta_1,\eta_2,\dots,\eta_\leng$ are arbitrary parameters
controlling the ``inhomogeneity'' of the model.
The pre\-factor $q^{-\frac{N}{2}}$ has been introduced
for convenience, to
ensure the normalization condition~\eqref{normT} below.
With the following graphical representation for the $R$-matrix
\begin{gather*}
{\begin{tikzpicture}[baseline=(current bounding box.center)]
\node at (0,0) { \large $R_{12}(q\zeta/\eta)^{b_1 b_2}_{a_1 a_2}\ = \ $};
\draw[line width=0.04cm] (2.3,0) -- (4.7,0);
\draw[line width=0.04cm] (3.5,1.2) -- (3.5,-1.2);
\node at (3.8,1) {$b_2$};
\node at (3.8,-1.0) {$a_2$};
\node at (2.6,-0.35) {$a_1$};
\node at (4.7,-0.30) {$b_1$};
\draw [thick] ([shift=(-180:0.4cm)]3.5,0) arc (-180:-90:0.4cm);
\node at (2.0,-0.) {$\zeta$};
\node at (3.5,-1.6) {$\eta$};
\end{tikzpicture}
}
\end{gather*}
the transfer matrix is depicted in Fig.~\ref{pic2}.

The transfer matrix is an operator acting in the
``quantum'' space, which is formed by the direct product of $N$ two-dimensional
 spaces
\begin{gather}\label{vec1}
{\mathscr V}_N=\mathbb{C}^2_N\otimes
 \mathbb{C}^2_{N-1}\otimes\cdots\otimes\mathbb{C}^2_1.
\end{gather}
Its diagonalization problem was solved by Baxter~\cite{Baxter:1971cs}
via the coordinate Bethe ansatz. The eigenvectors
with $M$ down spins have the form
\begin{gather}\label{cba}
\boldsymbol{\Psi}=
\sum_{1\leq x_1<x_2<\cdots<x_M\leq N}\Psi(x_1,\dots,x_M) \sigma_{x_M}^-\cdots
\sigma_{x_1}^- \mid 0\rangle,
\end{gather}
where the (unnormalized) Bethe ansatz wave function reads as
\begin{gather}\label{wave}
\Psi(x_1,\dots,x_M)=\sum_{\hat P}A_{\hat P} \prod_{m=1}^M \phi_{{\hat P}m}(x_m)
\end{gather}
with suitably chosen coefficients $A_{\hat P}$ and functions $\phi_m(x)$.
The vector $|0\rangle$ denotes the state
with all spins up,
\begin{gather}\label{iaususa}
|\,0\,\rangle=
\underbrace{|\uparrow\rangle\otimes |\uparrow\rangle\otimes \cdots\otimes |\uparrow\rangle}_{N},
\end{gather}
and $\sigma^z_J$, $\sigma^\pm_J\equiv \frac{1}{2}\big(\sigma^x_J\pm\ri\sigma^y_J\big)$
stand for the Pauli matrices
that act in the $J$-th factor of the tensor product~\eqref{vec1}.
The summation is taken over all $M!$ permutations ${\hat P}$ of the
integers $(1,2,\dots,M)$. Note that the
number of down spins $M$ is simply related to the
eigenvalue $S^z$ of the $z$-component of the total spin operator
\begin{gather}
{\mathbb S}^z=\frac12 \sum_{J=1}^N\sigma^z_J\colon\qquad
{\mathbb S}^z{\boldsymbol \Psi} =S^z {\boldsymbol \Psi}, \qquad M=\frac12 N-S^z,
\end{gather}
which commutes with the transfer matrix.

To within an overall normalization of the wave function~\eqref{wave}
the results of~\cite{Baxter:1971cs} can be
summarized as follows:
\begin{gather}\label{ap}
A_{\hat P}=\prod_{1\le j < m\le M}
\frac{q\zeta_{\hat Pj}-q^{-1}\zeta_{\hat Pm}}
{\zeta_{\hat Pj}-\zeta_{\hat Pm}},
\end{gather}
where the set of complex numbers $\{\zeta_j\}_{j=1}^M$ satisfies
the system of algebraic equations
\begin{gather}\label{bae}
\prod_{J=1}^{N}\frac{\eta_J+q^{+1}\zeta_m}
{\eta_J+q^{-1}\zeta_m }
=-\omega^2q^{2S^z}\prod_{j=1}^M\frac{\zeta_j-q^{+2}\zeta_m}
{\zeta_j-q^{-2}\zeta_m},\qquad
m=1,2,\dots,M,
\end{gather}
and the functions $\phi_m(x)$ in~\eqref{wave} have the form
\begin{gather}\label{single}
\phi_m(x)=-\ri q^{\frac{1}{2}}\omega^{-1}
\frac{\big(q-q^{-1}\big)\zeta_m}{q\eta_x+\zeta_m}
\prod_{J=1}^{x-1}\frac{\eta_J+q\zeta_m}{q\eta_J+\zeta_m}.
\end{gather}
The corresponding eigenvalue of the transfer matrix,
\begin{gather*}
{\mathbb T}(\zeta) \boldsymbol{\Psi} =T(\zeta) \boldsymbol{\Psi},
\end{gather*}
is given by\footnote{Note that in the staggered case when $N$ is even and
$\eta_1=\eta_3=\cdots=\eta_{N-1}$, $\eta_2=\eta_4=\cdots =\eta_N$
the diagonalization problem for $\mathbb{T}(\zeta)$ is significantly
simplified. It becomes equivalent to that of the transfer
matrix of the homogeneous six-vertex model on the $45^\circ$-rotated
square lattice. The Bethe ansatz for this problem was worked
out by Baxter in an unpublished work~\cite{Baxter:1970unpb}.}
\begin{gather}
T(\zeta)=\omega^{+1}q^{+S^z}
\bigg(\prod_{J=1}^N \big(1+q^{-1}\zeta/\eta_J\big)\bigg)
\prod_{m=1}^M\frac{\zeta_m-q^{+2}\zeta}{\zeta_m-\zeta}\nonumber
\\ \hphantom{T(\zeta)=}
{}+\omega^{-1}q^{-S^z}\bigg(\prod_{J=1}^N \big(1+q^{+1}\zeta/\eta_J\big)\bigg)
\prod_{m=1}^M\frac{\zeta_m-q^{-2}\zeta}{\zeta_m-\zeta}.
\label{teigen}
\end{gather}
It is easy to see from the definition~\eqref{tmat} that the transfer matrix
is an $N$-th order polynomial in~the variable~$\zeta$ satisfying the conditions
\begin{gather}
\mathbb{T}(0)=\omega^{+1}q^{+\mathbb{S}^z}+\omega^{-1}q^{-\mathbb{S}^z}, \nonumber
\\
\lim_{\zeta\to\infty}\zeta^{-N}\mathbb{T}(\zeta)=
\big(\omega^{+1}q^{-\mathbb{S}^z}+\omega^{-1}q^{+\mathbb{S}^z}\big)
\prod_{J=1}^N\eta_J^{-1}.
\label{normT}
\end{gather}
Since a multiplication of all the $\eta_J$ by the same factor can be absorbed into
a redefinition of the spectral parameter $\zeta$, we will always assume
\begin{gather}\label{cond1}
\prod_{J=1}^N\eta_J=1.
\end{gather}

The Bethe state~\eqref{cba} may also be constructed
within the framework of the QISM~\cite{Faddeev:1979gh,Takhtajan:1979iv}.
Introduce the so-called monodromy matrix
\begin{gather}\label{M-in}
{\bm{M}}(\zeta)\!\equiv\!{\bm M}\big(\zeta\mid\eta_N,\eta_{N-1},\dots,\eta_1\big)
\!=\!q^{-\frac{N}{2}}\bm{R}_{N}\big(q\zeta/\eta_N\big)
\bm{R}_{N-1}\big(q\zeta/\eta_{N-1}\big)\cdots
\bm{R}_{1}\big(q\zeta/\eta_{1}\big),\!\!\!
\end{gather}
where $\bm{R}_{m}$ is given by~\eqref{rmat}, but regarded as
a two by two matrix
\begin{gather}\label{rmat2}
\bm{R}_m(q\zeta)=
\begin{pmatrix}
q^{\frac{1}{2}(1+\sigma_m^z)}+q^{\frac{1}{2}(1-\sigma_m^z)}\zeta
&-\big(q-q^{-1}\big)q \zeta\sigma^-_m
\\[1ex]
\big(q-q^{-1}\big)\sigma^+_m
& q^{\frac{1}{2}(1-\sigma_m^z)}+q^{\frac{1}{2}(1+\sigma_m^z)}\zeta
\end{pmatrix}\!,
\end{gather}
whose elements act
in the $m$-th factor of the tensor product~\eqref{vec1}.
It is convenient to denote the entries of the monodromy matrix as
\begin{gather}\label{Mmat}
\bm{M}(\zeta)=
\begin{pmatrix}
\hat{{\mathsf A}}(\zeta)
&a(\zeta)\hat{\mathsf B} (\zeta)
\\[1ex]
d(\zeta)\hat{\mathsf C}(\zeta)
& \hat{\mathsf D}(\zeta)
\end{pmatrix}\!,
\end{gather}
where $\hat{{\mathsf A}}$, $\hat{\mathsf B}$, $\hat{\mathsf C}$, $\hat{\mathsf D}$
are operators acting in ${\mathscr V}_N=\mathbb{C}^2_N\otimes
\mathbb{C}^2_{N-1}\otimes\cdots\otimes\mathbb{C}^2_1$,
while the functions
\begin{gather}\label{addef1}
a(\zeta)=-\ri\omega^{+1} q^{+\frac{N+1}{2}}\prod_{J=1}^N\big(1+q^{-1}\zeta/\eta_J\big),\qquad
d(\zeta)=+\ri\omega^{-1} q^{-\frac{N+1}{2}}\prod_{J=1}^N\big(1+q\zeta/\eta_J\big).
\end{gather}
With the above notations, the transfer matrix~\eqref{tmat}
is given by
\begin{gather}\label{Tmat1}
\mathbb{T}(\zeta)={\rm Tr}\big[\omega^{\sigma^z}\bm{M}(\zeta)\big]=
\omega^{+1} \hat{{\mathsf A}}(\zeta) +\omega^{-1} \hat{\mathsf D}(\zeta),
\end{gather}
while
the Bethe state~\eqref{cba}--\eqref{single},
corresponding to the set $\{\zeta_j\}_{j=1}^M$ solving the Bethe ansatz equations, is expressed as
\begin{gather}\label{Bstate1}
\boldsymbol{\Psi}=\hat{\mathsf B}(\zeta_M)\cdots
\hat{\mathsf B}(\zeta_2) \hat{\mathsf B}(\zeta_1)\,|\,0\,\rangle.
\end{gather}
Note that the following remarkable formula holds true:
\begin{gather}\label{asioisisa}
\hat{\mathsf C}(\zeta_1)\hat{\mathsf C} (\zeta_2)\cdots \hat{\mathsf C}(\zeta_M) \boldsymbol{\Psi}={\mathfrak K}[{\boldsymbol \Psi}]\, |\,0\,\rangle,
\end{gather}
where we use the notation
\begin{gather}
\mathfrak{K}[{\boldsymbol \Psi}]=
\big(q-q^{-1}\big)^{2M}\prod_{m\ne j}^{M}
\frac{q\zeta_j-q^{-1}\zeta_m}{\zeta_m-\zeta_j}\nonumber
{\rm det}\bigg[\delta_{j,m}\bigg(\kappa(\zeta_j)+\sum_{l=1}^{M}
\frac{(q+q^{-1})\zeta_j\zeta_{l}}
{(q\zeta_l-q^{-1}\zeta_j)(q\zeta_j-q^{-1}\zeta_l)}\bigg)
\\ \hphantom{\mathfrak{K}[{\boldsymbol \Psi}]=\big(q-q^{-1}\big)^{2M}\prod_{m\ne j}^{M}
\frac{q\zeta_j-q^{-1}\zeta_m}{\zeta_m-\zeta_j}{\rm det}\bigg[}
{}-\frac{\big(q+q^{-1}\big)\zeta_j\zeta_{m}}
{\big(q\zeta_m-q^{-1}\zeta_j\big)(q\zeta_j-q^{-1}\zeta_m)}\bigg]
\label{FinalNorm}
\end{gather}
and
\begin{eqnarray}
{\kappa}(\zeta)=-\sum_{J=1}^N
\frac{\zeta}{
\eta_J(1+q^{-1}\zeta/\eta_J)(1+q^{+1}\zeta/\eta_J)}.\nonumber
\end{eqnarray}
This formula with all $\eta_J=1$ was originally conjectured
by Gaudin, McCoy and Wu in~\cite{Gaudin:1981cyg}. Its~generalization and proof for the inhomogeneous
 case was worked out by Korepin in~\cite{Korepin:1982gg}.

\section{{\bf Q}-operators\label{sec3}}

The eigenvalues
of the transfer matrix
are polynomials in the variable~$\zeta$. Indeed, its matrix elements defined through~\eqref{tmat} and~\eqref{rmat} are polynomials of $\zeta$ and
its eigenvectors~\eqref{cba}--\eqref{single} do not depend on this variable.
The latter also implies that the matrices ${\mathbb T}(\zeta)$ with different
values of $\zeta$ mutually commute.
It turns out that
the model possesses a larger commuting family
which, together with the transfer matrix, contains
other important members~-- the so-called
Baxter ${ Q}$-operators~\cite{Baxter:1972hz}. This section gives
an overview of their construction as well as their main properties.

First note that, introducing the functions
\begin{gather}\label{Qdef1A}
A_+(\zeta)=\prod_{m=1}^M
\big(1-\zeta/\zeta_m\big),\qquad
f(\zeta)=\prod_{J=1}^N \big(1+\zeta/\eta_J\big),
\end{gather}
one can rewrite the expression for the eigenvalues $T(\zeta)$
\eqref{teigen} in the form
\begin{gather}\label{tqeig}
T(\zeta)A_+(\zeta)= \omega^{+1} q^{+S^z}f\big(q^{-1}\zeta \big)A_+\big(q^{+2}\zeta \big)
 + \omega^{-1} q^{-S^z}f\big(q^{+1}\zeta \big)A_+\big(q^{-2}\zeta \big).
\end{gather}
Since $T(\zeta)$ is a polynomial, the r.h.s.\ of
the last relation must vanish
at the zeroes of $A_+(\zeta)$, i.e., when $\zeta=\zeta_m$, so that
\begin{gather*}
\omega^{+1}q^{+S^z}f\big(q^{-1}\zeta_m \big)A_+\big(q^{+2}\zeta_m \big)
+ \omega^{-1}q^{-S^z}f\big(q^{+1}\zeta_m \big)A_+\big(q^{-2}\zeta_m\big)=0,\quad\
m=1,2,\dots,M.
\end{gather*}
These are precisely the Bethe ansatz equations~\eqref{bae}.
The functional relation~\eqref{tqeig} implies the existence of an operator ${\mathbb
 A}_+(\zeta)$, with eigenvalues $A_+(\zeta)$~\eqref{Qdef1A},
that commutes with the transfer matrix
${\mathbb T}(\zeta)$ and satisfies the operator version of~\eqref{tqeig}.\footnote{In this paper we use
blackboard bold symbols such as ${\mathbb T}$, ${\mathbb A}_\pm$, ${\mathbb Q}_\pm$,
${\mathbb S}^z$, \dots to denote the mutually commuting operators
acting in the space ${\mathscr V}_N$~\eqref{vec1}. For the operators acting in the same space which do not belong to the commuting family, say the entries
 ${\hat{{\mathsf A}}}$, ${\hat{\mathsf B}}$, ${\hat{\mathsf C}}$, ${\hat{\mathsf D}}$
 of the monodromy matrix~\eqref{Mmat}, the ``hat'' notation is employed.}
Actually there are two such
operators, which we denote as ${\mathbb
 A}_\pm(\zeta)$, that satisfy the operator relations of this
type
\begin{gather}\label{tqrel}
{\mathbb T}(\zeta){\mathbb A}_\pm(\zeta)=q^{\pm2{\mathbb P}}
f\big(q^{-1}\zeta \big){\mathbb A}_\pm\big(q^{+2}\zeta \big)
+q^{\mp2{\mathbb P}}f\big(q^{+1}\zeta \big){\mathbb A}_\pm\big(q^{-2}\zeta \big),
\end{gather}
where for convenience we have used the notation
\begin{gather}
\label{q2p-def}
q^{2{\mathbb P}}=\omega q^{\mathbb{S}^z}.
\end{gather}
Note that equation~\eqref{tqrel} can be rewritten in a way
so that it has the same form for the
``$+$'' and ``$-$'' cases. Introducing the operators
$\mathbb{Q}_\pm(\zeta)$,
which
differ from ${\mathbb A}_\pm(\zeta)$ by simple factors involving
 fractional powers of $\zeta$,
\begin{gather*}
 {\mathbb Q}_\pm(\zeta)=
 \zeta^{\pm {\mathbb P}} {\mathbb A}_\pm(\zeta),
\end{gather*}
one has
\begin{gather*}
 {\mathbb T}(\zeta){\mathbb Q}_\pm(\zeta)=f\big(q^{-1}\zeta \big){\mathbb Q}_\pm\big(q^{+2}\zeta \big)
 +f\big(q^{+1}\zeta \big){\mathbb Q}_\pm\big(q^{-2}\zeta \big)
 . 
\end{gather*}
The latter is the famous Baxter ``$TQ$-relation''~\cite{Baxter:1972hz}.
We prefer to work with the operators ${\mathbb A}_\pm(\zeta)$
as their matrix elements and eigenvalues are polynomials in $\zeta$.

Similar to $\mathbb{T}(\zeta)$, which can be expressed as
a trace of a $2\times 2$ matrix~\eqref{Tmat1},
the operators~${\mathbb A}_\pm(\zeta)$ are also
constructed as traces of certain monodromy matrices.
However, this time the trace is taken over
infinite dimensional representations of the so-called $q$-oscillator algebra.
The latter arises as an evaluation
representation of the Borel subalgebra of the quantum affine
algebra~$U_q\big(\widehat{\mathfrak{sl}}_2\big)$. The required $R$-matrices
(see~\eqref{Rpm} below) are obtained by
a suitable specialization of the universal
$R$-matrix~\cite{Khoroshkin:1991}.
Originally these calculations were performed
 in the context of the quantum KdV theory
\cite{Bazhanov:1996dr,Bazhanov:1998dq}, but the
same procedure can be readily applied to the inhomogeneous
six-vertex model.

The $q$-oscillator algebra is generated by the elements $\Hcal$ and
$\Ecal_\pm$ satisfying the commutation relations
\begin{gather}\label{qosc}
[\Hcal,\Ecal_\pm]=\pm2\Ecal_\pm,\qquad
q \Ecal_+\Ecal_--
q^{-1}\Ecal_-\Ecal_+=\frac{1}{q-q^{-1}} .
\end{gather}
The basic building blocks for the construction of $\mathbb{A}_\pm$
are the $2\times 2$ matrices
\begin{gather}
 {\mathbfcal { R}}^{(\pm)}_m(\zeta)=
 q^{\pm\frac{{\cal H}\otimes\sigma^z_m}{2}}+ \big(q-q^{-1}\big)
\big({\cal E}_\mp \otimes \sigma^+_m-\zeta {\cal E}_\pm\otimes \sigma^-_m\big) \nonumber
\\ \hphantom{{\mathbfcal { R}}^{(\pm)}_m(\zeta)=}
{}+\zeta\big( q-q^{-1}\big)q^{-\frac{{\cal H}}{2}}\big({\cal E}_+{\cal E}_--{\cal E}_-{\cal E}_+\big)
\otimes\sigma^\mp_m\sigma^\pm_m ,
\label{Rpm}
\end{gather}
which act in the $m$-th component of the tensor product~\eqref{vec1} and whose entries
involve the formal generators~\eqref{qosc}.
Let $\rho_\pm$ be representations of the $q$-oscillator algebra such that
the traces
\begin{gather*}
{\rm Tr}_{\rho_\pm}\big[q^{\pm 2{P}{\cal H}}\big]\ne 0 \qquad
{\rm with} \qquad \big|q^{2P}\big|>1
\end{gather*}
exist and are non-vanishing.
Then, using the notation~\eqref{q2p-def} and
assuming that the twist parameter $\omega$ is such that
\begin{gather}\label{omegacond}
|\omega|>\max\big(\big|q^{N/2}\big|,\big|q^{-N/2}\big|\big)
\end{gather}
one can introduce the operators
\begin{gather}\label{zpm1}
{\mathbb Z}_\pm={\rm Tr}_{\rho_\pm}\big[q^{\pm 2{\mathbb P}{\cal H}} \big]
\end{gather}
that act in the quantum space ${\mathscr V}_N$~\eqref{vec1}.
With these preparations, define
\begin{gather}\label{Adef}
{\mathbb A}_\pm(\zeta)={\mathbb Z}^{-1}_\pm
{\rm Tr}_{\rho_\pm}\big[\omega^{\pm\Hcal}
{\mathbfcal { R}}^{(\pm)}_N(\zeta/\eta_N)
{\mathbfcal { R}}^{(\pm)}_{N-1}(\zeta/\eta_{N-1}) \cdots
{\mathbfcal { R}}^{(\pm)}_1(\zeta/\eta_1)\big].
\end{gather}
It is easy to see that such operators commute with ${\mathbb S}^z$
and, therefore, with $\mathbb{Z}_\pm$.
As was first pointed out in~\cite{Bazhanov:1996dr}, the (normalized)
trace in~\eqref{Adef} is completely determined
by the commutation relations~\eqref{qosc} and the cyclic property of the
trace, so that the specific choice of the representations
$\rho_\pm$ is not
significant as long as the traces~\eqref{zpm1} exist and are non-vanishing.
This way it is possible to define
 ${\mathbb A}_\pm(\zeta)\in{\rm End}(\mathscr{V}_N)$
for all complex $\omega$ (except some discrete set
of isolated points, see below) through analytic continuation,
despite that
the definition~\eqref{Adef} applies literally to the case
\eqref{omegacond} only.

Following the line of~\cite{Bazhanov:1998dq},
one can show that the operators $\mathbb{A}_\pm(\zeta)$ commute with
the transfer matrix and among themselves for different values of the
spectral parameter:
\begin{gather*}
\big[\mathbb{A}_\pm(\zeta_1),{\mathbb{T}}(\zeta_2)\big]=
\big[\mathbb{A}_\pm(\zeta_1),\mathbb{A}_\pm(\zeta_2)\big]
=\big[\mathbb{A}_\pm(\zeta_1),\mathbb{A}_\mp(\zeta_2)\big]=0.
\end{gather*}
Furthermore they
satisfy a number of important operator valued relations,
which are derived algebraically from the decomposition properties
of products of representations of the $q$-oscillator algebra.
This includes the quantum Wronskian relation
\begin{gather}\label{qwron}
\big(q^{+2\mathbb P}-q^{-2\mathbb P}\big) f(\zeta)=
q^{+2\mathbb P}\mathbb{A}_+\big(q^{+1}\zeta\big)\mathbb{A}_-\big(q^{-1}\zeta \big)-
q^{-2\mathbb P} \mathbb{A}_-\big(q^{+1}\zeta\big)\mathbb{A}_+\big(q^{-1}\zeta \big)
 \end{gather}
{\samepage
as well as
\begin{gather*}
\big(q^{+2\mathbb P}-q^{-2\mathbb P}\big) {\mathbb T}(\zeta)=
q^{+4\mathbb P} \mathbb{A}_+\big(
q^{+2}\zeta\big)\mathbb{A}_-\big(q^{-2}\zeta \big)- q^{-4\mathbb P}
 \mathbb{A}_-\big(q^{+2}\zeta \big)\mathbb{A}_+\big(q^{-2}\zeta \big).
\end{gather*}
As a simple corollary of
the above two formulae, the operators $\mathbb{A}_\pm$
satisfy~\eqref{tqrel}.

}

From the definition~\eqref{Adef} it follows that the
eigenvalues of $\mathbb{A}_\pm(\zeta)$ are polynomials\vspace{-1ex}
\begin{gather}\label{qpmeig}
A_\pm(\zeta)=\prod_{m=1}^{M_\pm}\big(1-\zeta/\zeta_m^{(\pm)}\big)
\end{gather}
of degree\vspace{-1ex}
\begin{gather*}
M_\pm=\frac12 \leng\mp S^z.
\end{gather*}
Their roots $\zeta_m^{(\pm)}$ are determined by the Bethe ansatz
equations\vspace{-1ex}
\begin{gather*}
\frac{f\big(\zeta_m^{(\pm)} q^{+1}\big)}{f\big(\zeta_m^{(\pm)} q^{-1}\big)}
= -\omega^{\pm2}q^{\pm2S^z}\frac{{A}_\pm\big(\zeta_m^{(\pm)} q^{+2}\big)}
{{A}_\pm\big(\zeta_m^{(\pm)} q^{-2}\big)},\qquad
m=1,2,\dots,M_\p).
\end{gather*}
Formula~\eqref{qpmeig} implies that\vspace{-1ex}
\begin{gather}\label{Qasymp}
{\mathbb A}_\pm(\zeta)\to
\begin{cases}
1 &{\rm as}\quad \zeta\to 0,
\\
\zeta^{M_\pm} {\mathbb{A}}^{(\infty)}_\pm &{\rm as}\quad \zeta\to+\infty,
\end{cases}
\end{gather}
where the operators ${ \mathbb{A}}^{(\infty)}_\pm $ act in
${\mathscr V}_N$~\eqref{vec1} and their eigenvalues on the Bethe state~\eqref{Bstate1} read~as\vspace{-1ex}
\begin{gather*}
{A}^{(\infty)}_\pm=\prod_{j=1}^{M_\pm}\big({-}\zeta^{(\pm)}\big)^{-1}.
\end{gather*}
Combining~\eqref{Qasymp} with~\eqref{qwron} one obtains\vspace{-1ex}
\begin{gather*}
{ \mathbb{A}}^{(\infty)}_+{ \mathbb{A}}^{(\infty)}_-
=\frac{1-\omega^{2}q^{2\mathbb{S}^z}}{q^{2\mathbb{S}^z}-\omega^{2}}.
\end{gather*}

It should be pointed out that, defined through equation~\eqref{Adef},
the operators $\mathbb{A}_\pm(\zeta)$ obey the normalization condition
$\mathbb{A}_\pm(0)=\hat{\bf 1}$. However, in
the subspace with fixed $S^z$ and when
$\omega=\pm q^{-S^z+m}$
with $m=\pm1,\pm2,\dots$
the Bethe roots $\{\zeta_m\}$ for
some of the eigenvalues of $\mathbb{A}_\pm$
may become zero.
In this case, as it follows from~\eqref{qpmeig},
the normalization imposed by~\eqref{Adef} is not
suitable and that formula requires modification.
As our considerations are not sensitive to this subtlety,
we will continue to use~\eqref{Adef}
and assume that~\eqref{qpmeig} always holds.

Finally note that for $S^z=0$ both
$\mathbb{A}_\pm(\zeta)$ tend to the same operator
$\mathbb{A}(\zeta)$ as $\omega \to1$.
A careful taking of this limit in equation~\eqref{Adef} leads
 to a generalization of Baxter's formula
for the matrix elements of the $Q$-operator of the zero-field
six-vertex model in the sector with $S^z=0$:
\begin{gather}\label{Q0field}
\big(\mathbb{A}(\zeta)\big)_{a_N,a_{N-1},\dots,a_1}^{b_N,b_{N-1},\dots,b_1}
=q^{\frac{1}{4}\sum_{1\le J< K\le N}(a_K b_J -a_J b_K)}
\prod_{J=1}^N \big({-}\zeta/\eta_J\big)^{\frac{1}{4}(1-a_J)(1+ b_J)}.
\end{gather}
In the homogeneous case, when all the $\eta_J=1$, the last formula reduces
to equation~(101) of~\cite{Baxter:1972wg}.
Note that the
operator~\eqref{Q0field} is still normalized as
$\mathbb{A}(0)=1$, which may not be immediately obvious.

\section[C, P and T conjugations]{${\cal C}$, ${\cal P}$ and ${\cal T}$ conjugations}\label{sec4}

So far we have briefly reviewed the formal
algebraic aspects of the diagonalization problem for the commuting family
of operators
in the general inhomogeneous six-vertex model. All the parameters
$\{\eta_J\}$, $q$ and $\omega$ were assumed to be generic complex
numbers. Though the con\-s\-t\-raint~\eqref{cond1} was imposed,
it does not reduce the generality, as it can always be achieved via a~redefinition of the spectral parameter $\zeta$.
In this section we introduce three global conjugations, which are similar to
the charge conjugation ${\cal C}$, parity inversion ${\cal P}$,
and time reversal transformation ${\cal T}$ from quantum field theory.
We describe the restrictions on the parameters such that the
global involutions are consistent with the integrable structure, in the
sense that they preserve the family of commuting operators.

Let us first consider parity inversion.
This transformation may be defined via the following adjoint action on the
local spin operators
\begin{gather*}
\hat{\cal P}\sigma^{A}_J\hat{\cal P} =(\eta_{\leng+1-J})^{+\frac{1}{2}\sigma_{\leng+1-J}^z}\sigma^{A}_{\leng+1-J}
(\eta_{\leng+1-J})^{-\frac{1}{2}\sigma_{\leng+1-J}^z},\qquad
A=x,y,z,\quad J=1,\dots, \leng,
\end{gather*}
supplemented by the condition
\begin{gather}\label{oapsdop10242093}
\hat{\cal P}\,|\,0\,\rangle=|\,0\,\rangle,
\end{gather}
where $|\,0\,\rangle$ stands for the pseudovacuum~\eqref{iaususa}.\footnote{The definition of the generator $\hat{\cal P}$ contains an ambiguity in
the choice of the overall sign $\hat{\cal P}\mapsto c_N\hat{\cal P}$
with $c_N^2=1$. In this work it is fixed by the condition~\eqref{oapsdop10242093}.
In~\cite{Bazhanov:2020} the sign factor is chosen differently. The same also applies to
the generator of the charge conjugation $\hat{\cal C}$~\eqref{Cconj1}.}
Its matrix elements read as
\begin{gather}\label{oaisodi899832}
(\hat{\cal P})^{b_Nb_{N-1}\cdots b_1}_{a_Na_{N-1}\cdots a_1}=
\delta_{a_N}^{b_1}\delta_{a_{N-1}}^{b_2}\cdots \delta_{a_1}^{b_N} \prod_{J=1}^N\eta_J^{a_J/2}.
\end{gather}
In order for $\hat{\cal P}^2=\hat{\bf 1}$, one must require that
\begin{gather}\label{alpha-sym}
\eta_J=(\eta_{N+1-J})^{-1},\qquad J=1,2,\dots,N.
\end{gather}
Then the action of the parity conjugation on the transfer matrix and $\mathbb{A}_\pm(\zeta)$ is given by
\begin{gather}
{\hat {\cal P}}{\mathbb T}(\zeta\,|\,{\omega}){\hat {\cal P}}
=\zeta^N{\mathbb T}\big(\zeta^{-1}\,|\,{\omega^{-1}}\big),\nonumber\\
{\hat {\cal P}}{\mathbb A}_\pm(\zeta\,|\,\omega){\hat {\cal P}}
=\zeta^{(\frac{N}{2}\mp\mathbb{S}^z)}{\mathbb A}_\pm\big(\zeta^{-1}\,|\,\omega^{-1}\big)
\big[{\mathbb A}^{({\infty})}_\pm\big(\omega^{-1}\big)\big]^{-1},
\label{Ptrans}
\end{gather}
where we explicitly indicate the dependence
on the twist parameter $\omega$.
These relations may be deduced directly from the definition of $\mathbb{T}(\zeta)$ and $\mathbb{A}_\pm(\zeta)$.
For instance, the first equality follows from~\eqref{Tmat1} and~\eqref{M-in}
as well as the simple property
\begin{gather*}
\big(\sigma^x\otimes \zeta^{\frac{1}{2} \sigma^z_m}\big) \big(\bm{R}_{m}(q\zeta)\big)^{t}
\big(\sigma^x\otimes \zeta^{\frac{1}{2}\sigma^z_m}\big)^{-1}=\zeta\bm{R}_{m}\big(q\zeta^{-1}\big),
\end{gather*}
where $\big(\bm{R}_{m}(q\zeta)\big)^{t}$
stands for the transpose of the $2\times 2 $ operator valued matrix~\eqref{rmat2}. Since
$\mathbb{T}(\zeta\,|\,\omega)$ and $\mathbb{A}_\pm(\zeta\,|\,\omega)$ do not commute amongst themselves for different values of the twist parameter $\omega$,
formula~\eqref{Ptrans} shows that unless $\omega^{2}=1$, the parity
inversion does not preserve the commuting family of operators.

The situation is similar for the charge conjugation.
Its generator may be introduced as
\begin{gather}\label{Cconj1}
\hat{\cal C}=\prod_{J=1}^N(\eta_J)^{\frac{1}{2}\sigma^z_J} \sigma^x_J,
\end{gather}
whose adjoint action on the local spin operators is given by
\begin{gather*}
{\hat {\cal C}} \sigma_J^\pm {\hat {\cal C}}= \eta_J^{\mp 1}\sigma_J^\mp,\qquad
 {\hat {\cal C}} \sigma_J^z {\hat {\cal C}} =-\sigma_J^z,\qquad J=1,\dots, \leng.
 \end{gather*}
For arbitrary complex values of the inhomogeneities
one has
\begin{eqnarray}\label{Ctrans}
{\hat {\cal C}} {\mathbb T}(\zeta\,|\,\omega) {\hat {\cal C}}
={\mathbb T}\big(\zeta\,|\,\omega^{-1}\big), \qquad
{\hat {\cal C}}{\mathbb A}_\pm(\zeta\,|\,\omega)
{\hat {\cal C}}={\mathbb A}_\mp\big(\zeta\,|\,\omega^{-1}\big).
\end{eqnarray}
The proof of the first equality is based on the relation
\begin{gather*}
\big(\sigma^x\otimes \zeta^{-\frac{1}{2} \sigma^z_m}\sigma^y_m\big) \bm{R}_{m}(\zeta)
\big(\sigma^x\otimes \zeta^{-\frac{1}{2} \sigma^z_m}\sigma^y_m\big)^{-1}=\bm{R}_{m}(\zeta).
\end{gather*}

Though
the ${\cal C}$ and ${\cal P}$ transformations do not respect the integrable structure themselves
for $\omega^2\ne 1$,
by combining them together
one gets
\begin{gather}
{\hat {\cal C}}{\hat {\cal P}}
{\mathbb T}(\zeta){ \hat {\cal C}}{\hat {\cal P}}=
\zeta^{N} {\mathbb T}\big(\zeta^{-1}\big),\nonumber
\\
{\hat {\cal C}}{\hat {\cal P}}
{\mathbb A}_\pm(\zeta){ \hat {\cal C}}{\hat {\cal P}}
=\zeta^{\frac{N}{2}\mp\mathbb{S}^z} {\mathbb A}_\mp\big(\zeta^{-1}\big)
\big[ \mathbb{A}^{(\infty)}_\pm\big]^{-1}
\label{CP}
\end{gather}
provided that the conditions~\eqref{alpha-sym} are imposed.
Here all the original and transformed operators
correspond to the same value of $\omega$ so that they commute with each other and
 have the same set of eigenvectors. This has an important consequence.
Since the ${\cal C}{\cal P}$ conjugation intertwines
the~sectors with $S^z$ and $-S^z$
it is always possible to focus on the case with $S^z\ge 0$.
Further, the second line of equation~\eqref{CP} implies
that for the construction of the Bethe states
\eqref{Bstate1} in~the~sector with arbitrary $S^z$
it is sufficient to consider the operator $\mathbb{A}_+(\zeta)$.
For these reasons we focus on~the zeroes of the eigenvalues of
$\mathbb{A}_+ $ and use the notation $\zeta_m\equiv \zeta_m^{(+)}$ and
$M\equiv M_+\le\frac{N}{2}$.

Up to this point there was no essential need to impose any reality conditions
on the parameters of the model. However, for the
time reversal transformation, the reality conditions become crucial.
Having in mind applications to the ${\cal Z}_2$ invariant model studied in
\cite{Jacobsen:2005xz,Ikhlef:2008zz,Ikhlef:2011ay,Frahm:2012eb,
Candu:2013fva,Frahm:2013cma,Bazhanov:2019xvy,Bazhanov:2020},
we will assume that $q$ and $\omega$ are unimodular
and parameterize them as
\begin{gather*}
q=\re^{\ri\gamma},\qquad
0<\gamma<\pi,
\end{gather*}
and
\begin{gather*}
\omega=\re^{\ri\pi{\tt k}},\qquad
-\frac12<{\tt k}\le \frac12.
\end{gather*}
The $R$-matrix then obeys the property
\begin{gather*}
(\sigma^x\otimes\sigma^x)\big({\bm R}(q\zeta)\big)^*
(\sigma^x\otimes\sigma^x)=q^{-1}\zeta^*{\bm R}(q/\zeta^*).
\end{gather*}
As for the inhomogeneities,
three cases will be considered:
\begin{enumerate}\itemsep=0pt
\item[$(i)$] All $\eta_J$ are unimodular complex numbers satisfying~\eqref{alpha-sym}:
\begin{gather}\label{case1a}
\eta_{J}^*=\eta_J^{-1}=\eta_{N+1-J}.
\end{gather}
\item[$(ii)$] All the inhomogeneities are real:
\begin{gather}\label{case2a}
\eta_{J}^*=\eta_J=\big(\eta_{N+1-J}\big)^{-1}.
\end{gather}
\item[$(iii)$] In the third case we take the number of columns in the lattice to be even
and divide the inhomogeneities into two groups: $\{\eta_J\}_{J=1}^{N/2}$ and
$\{\eta_{N/2+J}\}_{J=1}^{N/2}$.
 The first group $\{\eta_J\}_{J=1}^{N/2}$ are taken to have the same absolute value $\Lambda>0$,
 while for the other group $|\eta_{N/2+1}|=\cdots=|\eta_{N}|=\Lambda^{-1}$.
The arguments of $\eta_J$ and $\eta_{N/2+J}$ are chosen to be the same
\begin{gather}\label{case3}
\eta_J=\Lambda^{-1} \omega_J, \qquad
\eta_{N/2+J}=\Lambda\omega_J,\qquad J=1,2,\dots,N/2,
\end{gather}
where
$\{\omega_J\}_{J=1}^{N/2}$ are unimodular complex numbers.
Notice that the constraint~\eqref{alpha-sym}, which is always assumed,
implies
\begin{gather}\label{case3a}
\omega_J=(\omega_J^*)^{-1}=(\omega_{N/2+1-J})^{-1}.
\end{gather}
\end{enumerate}

The time reversal transformation
 is realized as an anti-unitary operator acting on
an arbitrary state ${\boldsymbol \psi}\in{\mathscr V}_N$ as
\begin{gather}\label{Tdef}
\hat{\cal T}{\boldsymbol \psi}=\hat{\mathsf U}{\boldsymbol \psi}^*,
\end{gather}
where the asterisk $(*)$ stands for complex conjugation.
The matrix $\hat{\mathsf U}$ satisfies the condition
\begin{gather}\label{Umatrel1}
\hat{{\mathsf U}}^*=\hat{{\mathsf U}}^{-1}
\end{gather}
which ensures that $\hat{\cal T}^2=\hat{\bf 1}$.
For the first case~\eqref{case1a} we set the matrix
$\hat{\mathsf U}$ to be
\begin{gather}\label{T1}
\eta_{J}^*=\eta_J^{-1}=\eta_{N+1-J}\colon\quad
\hat{\mathsf U}=\prod_{J=1}^N\sigma^x_J.
\end{gather}
Then it follows that
\begin{gather}
{\hat {\cal T}}{\mathbb T}(\zeta){ \hat {\cal T}}=(\zeta^*)^N
{\mathbb T}\big((\zeta^*)^{-1}\big),\nonumber
\\
{\hat {\cal T}}{\mathbb A}_\pm(\zeta){ \hat {\cal T}}
=(\zeta^*)^{\frac{N}{2}\mp\mathbb{S}^z}{\mathbb A}_\mp\big((\zeta^*)^{-1}\big)
\big[{\mathbb A}^{(\infty)}_\mp\big]^{-1},
\label{Ttrans}
\end{gather}
i.e., defined in this way the time reversal transformation
preserves the integrable structure of~the model similar to ${\cal C}{\cal P}$.
Combining all three transformations~\eqref{Ctrans},~\eqref{Ptrans} and
\eqref{Ttrans} together, yields
\begin{gather}\label{isasasu1}
{\hat {\cal C}}{\hat {\cal P}}{\hat {\cal T}}
{\mathbb A}_\pm(\zeta){\hat {\cal C}}{\hat {\cal P}}{\hat {\cal T}}
={\mathbb A}_\pm(\zeta^*),\qquad
{\hat {\cal C}}{\hat {\cal P}}{\hat {\cal T}}
{\mathbb T}(\zeta){\hat {\cal C}}{\hat {\cal P}}{ \hat {\cal T}}
={\mathbb T}(\zeta^*).
\end{gather}
For the Bethe states, the above equation
together with~\eqref{Qdef1A} implies that the sets $\{\zeta_m\}$ corresponding to
 $\boldsymbol{\Psi}$ and ${\cal \hat{C}\hat{P}\hat{T}}\boldsymbol \Psi$ are complex conjugate to each other. Notice that the phase assig\-n\-ment in~\eqref{single} and consequently in~\eqref{addef1}
has been chosen in such a way that
\begin{gather}\label{CPTBethe1}
{\cal \hat{C}\hat{P}\hat{T}}\boldsymbol{\Psi}(\{\zeta_j\})=\boldsymbol{\Psi}(\{\zeta_j^*\}).
\end{gather}

In the second case $(ii)$
it is also possible to introduce the time-reversal symmetry in such
a way that relations~\eqref{Ttrans} are preserved.
To proceed we will need to establish how the transfer matrix and,
more generally, the monodromy matrix $\bm{M}(\zeta\,|\,\eta_N,\eta_{N-1},\dots,\eta_1)$~\eqref{M-in},
behaves under an interchange of the inhomogeneities.
The eigenvalues of the transfer matrix~\eqref{teigen} depend on~$\{\eta_J\}$
both explicitly
and implicitly via the Bethe roots which satisfy equation~\eqref{bae}.
In both cases the inhomogeneities enter only through symmetric combinations,
so that the eigenvalues are symmetric functions of them.
 Therefore two transfer matrices obtained from each other by a~per\-mu\-ta\-tion of the same set of
inhomogeneities must be connected by a similarity transformation in the quantum space $\mathscr{V}_N$.
In fact, the same holds true for all the entries of the monodromy matrix as well.
The group of permutations of the ordered set $(\eta_J)$ is generated by
elementary permutations of neighbouring pairs:
\[
(\eta_N,\dots,\eta_{n+1},\eta_n,\dots,\eta_1)\mapsto
(\eta_N,\dots,\eta_{n},\eta_{n+1},\dots,\eta_1).
\]
The latter are realized through the relation
\begin{gather}
{\bm M}(\zeta\,|\,\eta_N,\dots,\eta_{n+1},\eta_n,\dots,\eta_{1}) \check{\bm{R}}_{n+1,n}(\eta_{n+1}/\eta_{n})\nonumber
\\ \qquad
=\check{\bm{R}}_{n+1,n}(\eta_{n+1}/\eta_{n}) {\bm M}(\zeta\,|\,\eta_N,\dots,\eta_{n},\eta_{n+1},\dots,\eta_{1}),
\label{YBE321}
\end{gather}
which is a simple consequence of the Yang--Baxter equation.
Here $\check{\bm{R}}_{n+1,n}$
is expressed in terms of the $R$-matrix~\eqref{rmat} as
\begin{gather}\label{Rcheckdef1}
\check{\bm{R}}_{n+1,n}(\zeta)=\frac{1}{q-q^{-1}\zeta} {\bm{R}}_{n+1,n}(-\zeta){\bm P}_{n+1,n}
\end{gather}
with ${\bm P}_{n+1,n}$ standing for the operator that permutes the $(n+1)$-th and $n$-th
factor in the tensor product~\eqref{vec1}.
For the operation that completely reverses the order
of the $\{\eta_J\}$ it is easy to see that
\begin{gather*}
{\bm M}(\zeta\,|\,\eta_N,\eta_{N-1},\dots,\eta_1)\mapsto
{\bm M}(\zeta\,|\,\eta_1,\eta_{2},\dots,\eta_{N})=\hat{\mathsf S}_2^{-1}{\bm M}
(\zeta\,|\,\eta_N,\eta_{N-1},\dots,\eta_1) \hat{\mathsf S}_2,
\end{gather*}
where the matrix
$\hat{\mathsf S}_2\in{\rm End}({\mathscr V}_N)$ is given by the ordered product
\begin{gather}\label{Sdef1b}
\hat{\mathsf S}_2=
\overset{\displaystyle \curvearrowleft}{\prod_{m=2}^{\leng}}
\Bigg[\,\overset{\displaystyle \curvearrowleft}{\prod_{n=N-m+1}^{\leng-1}}
\check{\bm{R}}_{n+1,n}\big(\eta_{m}/\eta_{n+m-N}\big)\Bigg].
\end{gather}
The time-reversal transformation corresponding to~\eqref{case2a}
 is defined through~\eqref{Tdef} with
\begin{gather*}
\eta_{J}^*=\eta_J=(\eta_{N+1-J})^{-1}\colon\quad
\hat{\mathsf U}=\hat{\mathsf S}_2\prod_{J=1}^N\sigma^x_J.
\end{gather*}
One can check that
the condition~\eqref{Umatrel1} is satisfied, and
that equations~\eqref{Ttrans},~\eqref{isasasu1} remain intact.

Finally let's turn to the last case $(iii)$, where the inhomogeneities are constrained by the conditions~\eqref{case3} and~\eqref{case3a}.
For this purpose we need to consider the similarity transformation in the quantum space
that interchanges the two
groups of inhomogeneities for the monodromy matrix, while preserving their ordering
inside of each group
\begin{gather*}
{\bm M}(\zeta\,|\,\eta_{N/2},\dots,\eta_{1},\eta_{N},\dots,\eta_{N/2+1})
=\hat{\mathsf S}_3^{-1} {\bm M}(\zeta\,|\,\eta_{N},\dots,\eta_{N/2+1},
\eta_{N},\dots,\eta_{1})\hat{\mathsf S}_3.
\end{gather*}
One can show that
\begin{gather}\label{Sdef3}
\hat{\mathsf S}_3=\overset{\displaystyle \curvearrowleft}{\prod_{m=N/2+1}^{N}}
\Bigg[\overset{\displaystyle \curvearrowleft}{\prod_{n=0}^{N/2-1}}
\check{\bm R}_{m-n,m-n-1}\big(\eta_{N-n}/\eta_{m-N/2}\big)\Bigg].
\end{gather}
Introduce the matrix $\hat{\mathsf U}$ as
\begin{gather*}
\eta_J=\Lambda^{-1} \omega_J, \qquad
\eta_{J+N/2}=\Lambda \omega_J, \qquad
|\omega_J|=1\colon\quad \hat{\mathsf U}=\hat{\mathsf S}_3\prod_{J=1}^N\sigma^x_J.
\end{gather*}
It turns out that it also satisfies~\eqref{Umatrel1} and,
 for the time-reversal conjugation~\eqref{Tdef} with
$\hat{\mathsf U}$ defined as above,
equations~\eqref{Ttrans} and~\eqref{isasasu1} continue to hold.

This way
in all of the cases $(i)$--$(iii)$ of the reality conditions imposed on the inhomogeneities,
it is possible to introduce the anti-unitary operator~\eqref{Tdef} such that
equations~\eqref{Ttrans} and~\eqref{isasasu1} are satisfied. The matrix
$\hat{\mathsf U}$ takes the general form
\begin{gather*}
\hat{\mathsf U}=\hat{\mathsf S}_{\cal T} \prod_{J=1}^N\sigma^x_J,
\end{gather*}
where $\hat{\mathsf S}_{\cal T}$ depends on the particular case being considered.
Namely when the inhomogeneities are constrained by~\eqref{case1a},
$\hat{\mathsf S}_{\cal T}$ is the identity operator.
For the cases~\eqref{case2a} and~\eqref{case3}
$\hat{\mathsf S}_{\cal T}$ is given by $\hat{\mathsf S}_2$~\eqref{Sdef1b} and
$\hat{\mathsf S}_3$~\eqref{Sdef3}, respectively.

\section{Hermitian structure\label{sec5}}

There are many ways to introduce the
Hermitian structure, i.e.,
the sesquilinear form for the states
and
the Hermitian conjugation of operators
in the $2^N$ dimensional linear space
$\mathscr{V}_N$~\eqref{vec1}.
Here we discuss
the Hermitian structures that are
consistent with the
integrable structure of the model.

Consider each factor in the product in the r.h.s.\ of
equation~\eqref{M-in}.
Under the standard matrix (``dagger'') conjugation
in the quantum space $\mathscr{V}_N$, they transform as
\begin{gather*}
\big[q^{-\frac{1}{2}} \bm{R}_m\big(q\zeta/\eta_m\big)\big]^\dag
=\sigma^x \big(\zeta^*/\eta_m^*\big)^{-\frac{1}{2}\sigma^z}
\big[q^{-\frac{1}{2}}\bm{R}_m\big(q\zeta^*/\eta_m^*\big)\big]
\big(\zeta^*/\eta_m^*\big)^{+\frac{1}{2}\sigma^z}\sigma^x.
\end{gather*}
In turn, taking the $\dag$-conjugation of the monodromy matrix results in
\begin{gather}
\label{eq42}
\big[\bm{M}(\zeta\,|\,\eta_N,\eta_{N-1},\dots,\eta_1)\big]^\dag=
\sigma^x \big(\zeta^*\big)^{-\frac{1}{2}\sigma^z}\hat{\mathsf{V}}
\bm{M}(\zeta^*\,|\,\eta_N^*,\eta_{N-1}^*,\dots,\eta_1^*)\hat{\mathsf{V}}^{-1}
\big(\zeta^*\big)^{+\frac{1}{2}\sigma^z}\sigma^x,
\end{gather}
where
\begin{gather*}
\hat{\mathsf{V}}=\prod_{J=1}^N (\eta_J^*)^{-\frac{1}{2}\sigma_J^z}.
\end{gather*}
Now suppose that the (non-ordered) set of inhomogeneities coincides with the complex conjugated set:
\begin{gather}\label{conj1}
\big\{\eta_J^*\big\}_{J=1}^N=\big\{\eta_J\big\}_{J=1}^N.
\end{gather}
In particular, this property holds for all the three cases~\eqref{case1a},
\eqref{case2a} and~\eqref{case3} considered above.
Then there exists a similarity transformation such that
\begin{gather}
\label{Sdef1a}
\bm{M}(\zeta\,|\,\eta_N,\eta_{N-1},\dots,\eta_1)=
\hat{\mathsf S}^{-1} \bm{M}(\zeta\,|\,\eta_N^*,\eta_{N-1}^*,\dots,\eta_1^*) \hat{\mathsf S},\qquad
 \hat{\mathsf S}\in{\rm End}(\mathscr{V}_N).
\end{gather}
Combining the latter with equation~\eqref{eq42}, yields
\begin{gather}
\label{aiasisau}
[\bm{M}(\zeta)]^\dag=\sigma^x(\zeta^*)^{-\frac{1}{2}\sigma^z}
\big[\hat{\mathsf X}\bm{M}(\zeta^*) \hat{\mathsf X}^{-1}\big]
\big(\zeta^*\big)^{+\frac{1}{2}\sigma^z}\sigma^x
\end{gather}
with the Hermitian matrix $\hat{\mathsf X}$ given by
\begin{gather}
\label{hhsaysys}
\hat{\mathsf X}=\hat{\mathsf X}^\dag=\hat{\mathsf V} \hat{\mathsf S}^{-1},\qquad
\hat{\mathsf X}\in{\rm End}(\mathscr{V}_N).
\end{gather}
Formula~\eqref{aiasisau} implies that, in general, the transfer matrix is
 not Hermitian w.r.t.~the~$\dag$-con\-ju\-ga\-tion. However, for an arbitrary
$\hat{\mathsf O}\in{\rm End}({\mathscr V}_N)$, one can
introduce the non-standard con\-ju\-ga\-tion~as
\begin{gather}
\label{ddag1}
\hat{\mathsf O}^\ddag=\hat{\mathsf X}^{-1}\hat{\mathsf O}^\dag \hat{\mathsf X}.
\end{gather}
Since the matrix $ \hat{{\mathsf X}}$ is Hermitian~\eqref{hhsaysys}, it is guaranteed
 that $\big(\hat{\mathsf O}^\ddag\big)^\ddag=\hat{\mathsf O}$ so that~\eqref{ddag1}
 indeed defines a~conjugation.
Then equation~\eqref{aiasisau} can be rewritten as the $\ddag$-con\-ju\-ga\-tion
condition for~the operator valued entries of the monodromy matrix~\eqref{Mmat}:
\begin{gather}
\big[{\hat{\mathsf A}}(\zeta)\big]^\ddag={\hat{\mathsf D}}(\zeta^*),\qquad
\big[{\hat{\mathsf D}}(\zeta)\big]^\ddag={\hat{\mathsf A}}(\zeta^*),\nonumber
\\
\big[{\hat{\mathsf B}}(\zeta)\big]^\ddag=\zeta^*{ \hat{\mathsf C}}(\zeta^*),\qquad
\big[{\hat{\mathsf C}}(\zeta)\big]^\ddag=(\zeta^*)^{-1}{ \hat{\mathsf B}}(\zeta^*).
\label{ABCDconj1}
\end{gather}
In turn, these relations imply
\begin{subequations}\label{ddagBCT}
\begin{gather}\label{ddagBCTa}
\big[\mathbb{T}(\zeta)\big]^\ddag=\mathbb{T}(\zeta^*).
\end{gather}
In a similar way, one can show that
\begin{gather}\label{ddagBCTb}
\big[\mathbb{A}_\pm(\zeta)\big]^\ddag=\mathbb{A}_\pm(\zeta^*).
\end{gather}
\end{subequations}

As a matter of fact, there is a more general version of~\eqref{ddag1}
for which the condition similar to~\eqref{ddagBCT} holds.
Indeed consider the conjugation of the form
\begin{gather}
\label{star1}
\hat{\mathsf O}^\star=\mathbb{Y}^{-1}\hat{\mathsf O}^\ddag\mathbb{Y}
=\mathbb{Y}^{-1}\hat{{\mathsf X}}^{-1}\hat{\mathsf O}^\dag \hat{{\mathsf X}}\mathbb{Y}.
\end{gather}
In order to fulfill the requirement $\big(\hat{\mathsf O}^\star\big)^\star=\hat{\mathsf O}$,
the operator $\mathbb{Y}$ should be such that
\begin{gather}
\label{Ycond1}
\mathbb{Y}^\ddag=\mathbb{Y}.
\end{gather}
If, in addition,
\begin{gather}
\label{Ycond2}
\big[\mathbb{Y},\mathbb{A}_\pm(\zeta)\big]=0,
\end{gather}
then it follows from~\eqref{ddagBCT} that
\begin{gather}
\label{starBCT}
\big[\mathbb{T}(\zeta)\big]^\star=\mathbb{T}(\zeta^*),\qquad
\big[\mathbb{A}_\pm(\zeta)\big]^\star=\mathbb{A}_\pm(\zeta^*).
\end{gather}
The $\ddag$-conjugation is a special case of the $\star$-conjugation with $\mathbb{Y}=\hat{\bf 1}$.

For a given ${\mathbb Y}\in{\rm End}({\mathscr V}_N)$ satisfying~\eqref{Ycond1}
and~\eqref{Ycond2} there is a unique
sesquilinear form $({\boldsymbol \psi}_2,{\boldsymbol \psi}_1)_{\star}$
$({\boldsymbol \psi}_{1,2}\in{\mathscr V}_N)$
 that is compatible with the $\star$-conjugation~\eqref{star1}. It is defined by the
requirement
\begin{gather*}
\big(\bm{\psi}_2,\hat{\mathsf O}\bm{\psi}_1\big)_{\star}=
\big(\hat{\mathsf O}^\star\bm{\psi}_2,\bm{\psi}_1\big)_{\star}
\end{gather*}
and the overall normalization condition for the pseudovacuum $\bm{\Psi}_0\equiv|\,0\,\rangle$~\eqref{iaususa}:
\begin{gather*}
\big(\bm{\Psi}_0,\bm{\Psi}_0\big)_{\star}=1.
\end{gather*}
The second equality in~\eqref{starBCT}, together with the fact that the spectrum of~the operators
$\mathbb{A}_\pm(\zeta)$ is expected to be non-degenerate for generic values of~${\tt k}$, implies that the Bethe states~\eqref{Bstate1} satisfy an orthogonality condition of the form
\begin{gather*}
\big({\boldsymbol \Psi}',{\boldsymbol \Psi} \big)_{\star}=0\qquad
{\rm unless}\qquad
{\boldsymbol \Psi}'={\hat {\cal C}}{\hat {\cal P}}{\hat {\cal T}}{\boldsymbol \Psi}.
\end{gather*}
Let us define the ``norm'' for an arbitrary state
${\boldsymbol\psi}\in{\mathscr V}_N$ as
\begin{gather*}
{\mathfrak N}_Y[{\boldsymbol \psi}]:=
\big({\hat {\cal C}}{\hat {\cal P}}{\hat {\cal T}}{\boldsymbol \psi}, {\boldsymbol \psi}\big)_{\star} .
\end{gather*}
For the Bethe state
corresponding to the set of Bethe roots $\{\zeta_m\}_{m=1}^M$, taking into account equations~\eqref{ABCDconj1} and~\eqref{asioisisa}, its norm is given by
\begin{gather}
\label{KYnorm1}
\mathfrak{N}_Y[{\boldsymbol \Psi}]=Y(\zeta_1,\dots,\zeta_M) \mathfrak{K}[{\boldsymbol \Psi}]
\prod_{m=1}^M\zeta_m,
\end{gather}
where $Y(\zeta_1,\dots,\zeta_M)$ stands for the eigenvalue of the operator ${\mathbb Y}$ on
the Bethe state ${\boldsymbol \Psi}$.

One can construct a variety of matrices
$\mathbb{Y}$ obeying equations~\eqref{Ycond1} and~\eqref{Ycond2}. For exam\-ple,
taking into account~\eqref{ddagBCTb}, it is easy to see that
both conditions are satisfied for~${\mathbb Y}=\re^{\ri\pi ({\mathbb S}^z-\frac{N}{2})} \mathbb{A}^{(\infty)}_+$,
where $\mathbb{A}_+^{(\infty)}$ is defined through the formula~\eqref{Qasymp}.
In this case the eigenvalues of $\mathbb{Y}$
are given by the product $\prod_{m=1}^M\zeta_m^{-1}$, so that the norm~\eqref{KYnorm1}
coincides with $\mathfrak{K}[\boldsymbol \Psi]$. In other words,
the functional formally introduced in~\eqref{FinalNorm} may be interpreted as
a particular norm of~the Bethe state
\begin{gather}\label{uassysay}
\mathfrak{K}[{\boldsymbol \Psi}]=\mathfrak{N}_Y[{\boldsymbol \Psi}]\qquad
{\rm with}\qquad
{\mathbb Y}=\re^{\ri\pi ({\mathbb S}^z-\frac{N}{2})}\mathbb{A}^{({\infty})}_+.
\end{gather}

It should be emphasized that different operators $\mathbb{Y}$ may lead to equivalent
Hermitian structures.
Replacing $\mathbb{Y}$ by ${\mathbb{Y}}'$ such that
\begin{gather}\label{normalize1}
{Y}'(\zeta_1,\dots,\zeta_M)=\alpha(\zeta_1,\dots,\zeta_M)
\big(\alpha(\zeta_1^*,\dots,\zeta_M^*)\big)^*Y(\zeta_1,\dots,\zeta_M)
\end{gather}
with $(\alpha(\zeta_1,\dots,\zeta_M))^{\pm 1}\ne0$ for \emph{any} Bethe state, may be compensated for
by a change of the normalization
\begin{gather*}
{\boldsymbol \Psi}\mapsto \alpha(\zeta_1,\dots,\zeta_M){\boldsymbol \Psi}.
\end{gather*}
However,
though the Hermitian structures corresponding to $\mathbb{Y}$ and $\mathbb{Y}'$~\eqref{normalize1} are
 equivalent in~the formal linear algebra sense, the action of the conjugation~\eqref{star1}
on the local spin operators could be radically different. In particular, a conjugation that acts
locally on~$\sigma_J^A$ may become highly non-local
upon the substitution $\mathbb{Y}\mapsto \mathbb{Y}'$.

The above discussion is valid for any set of inhomogeneities provided that~\eqref{conj1}
is satisfied. However, the Hermitian matrix $\hat{{\mathsf X}}$~\eqref{hhsaysys}, which was left unspecified,
depends on the precise form of the reality conditions imposed on $\{\eta_J\}$.
For the three cases from the previous section one has:
\begin{enumerate}\itemsep=0pt
\item[$(i)$] For unimodular inhomogeneities $\eta_J^*=\eta^{-1}_J=\eta_{N+1-J}$, the matrix
$\hat{\mathsf S}$~\eqref{Sdef1a} coincides with $\hat{\mathsf S}_2$ from~\eqref{Sdef1b}.
Hence $\hat{\mathsf X}$ is given by
\begin{gather}
\label{Xcase1}
\eta_{J}^*=\eta_J^{-1}=\eta_{N+1-J}\colon\quad
\hat{\mathsf X}=\bigg(\prod_{J=1}^N (\eta_J)^{\frac{1}{2}\sigma_J^z}\bigg)\hat{\mathsf S}_2^{-1}.
\end{gather}
\item[$(ii)$] In the case with $\eta_J$ real~\eqref{case2a}, the matrix $\hat{\mathsf S}$
 is the identity and $\hat{\mathsf X}$ reads as
\begin{gather}
\label{oiasd9821}
\eta_{J}^*=\eta_J=\big(\eta_{N+1-J}\big)^{-1}\colon\quad
\hat{\mathsf X}=\prod_{J=1}^N (\eta_J)^{-\frac{1}{2}\sigma_J^z}.
\end{gather}
\item[$(iii)$] For the case~\eqref{case3}, where the number of columns $N$ is assumed to be even,
the ordered set of inhomogeneities satisfies the condition
\begin{gather*}
(\eta_N^*,\dots, \eta_{N/2+1}^*,\eta_{N/2}^*,\dots,\eta_1^*)
=(\eta_{N/2+1},\dots, \eta_{N},\eta_{1},\dots,\eta_{N/2}).
\end{gather*}
It is convenient to write $\hat{\mathsf S}$ from~\eqref{Sdef1a} as the product
$\hat{\mathsf S}=\hat{\mathsf S}_{{\rm I}}\hat{\mathsf S}_{{\rm II}}$
and the matrices $\hat{\mathsf S}_{{\rm I}}$ and $\hat{\mathsf S}_{{\rm II}}$
reverse the order of the inhomogeneities of $\{\eta_J\}_{J=1}^{N/2}$ and
$\{\eta_{N/2+J}\}_{J=1}^{N/2}$
respectively, i.e.,
\begin{gather*}
{\bm M}(\zeta\,|\,\eta_{N},\dots,\eta_{N/2+1},\eta_{N/2},\dots,\eta_{1})=
\hat{\mathsf S}_{{\rm I}}^{-1}{\bm M}(\zeta\,|\,\eta_{N},\dots,\eta_{N/2+1},\eta_{1},\dots, \eta_{N/2})\hat{\mathsf S}_{\rm I},
\\
{\bm M}(\zeta\,|\,\eta_{N},\dots,\eta_{N/2+1},\eta_{N/2},\dots,\eta_{1})
=\hat{\mathsf S}_{\rm II}^{-1}{\bm M}(\zeta\,|\,\eta_{N/2+1},\dots,\eta_{N},\eta_{N/2}, \dots,\eta_{1})\hat{\mathsf S}_{\rm II}.
\end{gather*}
The expression for these matrices are obtained from equation~\eqref{Sdef1b}
by a simple replacement of~the indices
\begin{gather*}
\hat{\mathsf S}_{\rm I}=
\overset{\displaystyle \curvearrowleft}{\prod_{m=2}^{N/2}}
\Bigg[\,\overset{\displaystyle \curvearrowleft}{\prod_{n=N/2-m+1}^{N/2-1}}
\check{\bm{R}}_{n+1,n}\big(\eta_{m}/\eta_{n+m-N/2}\big)\Bigg],
 \\
\hat{\mathsf S}_{\rm II}=
\overset{\displaystyle \curvearrowleft}{\prod_{m=2}^{N/2}}
\Bigg[\,\overset{\displaystyle \curvearrowleft}{\prod_{n=N-m+1}^{N-1}}
\check{\bm{R}}_{n+1,n}\big(\eta_{m+N/2}/\eta_{n+m-N/2}\big)\Bigg].
\end{gather*}
Then the matrix $\hat{\mathsf{X}}$ is given by
\begin{gather*}
\eta_J=\Lambda^{-1} \omega_J,\qquad
\eta_{L+J}=\Lambda \omega_J,
\\
|\omega_J|=1\colon\quad
\hat{\mathsf{X}}=\bigg(\prod_{J=1}^{N/2} (\Lambda^{+1}\omega_J)^{+\frac{1}{2}\sigma_J^z} (\Lambda^{-1}\omega_J)^{+\frac{1}{2}\sigma_{J+N/2}^z}\bigg)
\big(\hat{\mathsf S}_{\rm I}\hat{\mathsf S}_{\rm II}\big)^{-1}.
\end{gather*}
\end{enumerate}

\section{Lattice translation symmetry\label{sec6}}
Translational invariance is a fundamental symmetry of a local
quantum field theory. Having in~mind the study of the universality classes
for the inhomogeneous six-vertex model, it is natural to focus
on the lattice, where the number of columns $N$ is divisible by
some integer $r$,
\begin{gather}
\label{aoisd981221}
N=r L,
\end{gather}
while the inhomogeneities satisfy the $r$-site periodicity conditions
\begin{gather}
\label{iisasausa}
\eta_{J+r}=\eta_{J} \qquad (J=1,2,\dots,N)
\end{gather}
(here by definition we take $\eta_{J+N}\equiv \eta_J$).
With these restrictions, the commuting family contains
a set of operators that play a central r\^ole for the
description of the critical behaviour.
The aim of this section is to introduce these operators and discuss
their basic properties.

\subsection[r-site translation operator]{$\boldsymbol r$-site translation operator}
The one-site translation operator may be defined
without any restrictions imposed on the parameters of the model.
Taking into account the quasi-periodic boundary conditions with twist parameter
$\omega=\re^{\ri\pi{\tt k}}$, its matrix elements are given by
\begin{gather}
\label{aiosd901221}
\big(\hat{{\cal K}}\big)_{a_{N} a_{N-1}\cdots a_1}^{b_{N}b_{N-1}\cdots b_1}
=\re^{\ri\pi{\tt k}a_1}\delta_{a_1}^{b_N}\delta_{a_N}^{b_{N-1}}\cdots\delta^{b_1}_{a_2}.
\end{gather}
The action of the one-site translation on the local
 spin operators reads as
\begin{gather*}
\hat{\cal K} \sigma_J^{A}\hat{\cal K}^{-1}= \sigma^{A}_{J+1}\qquad
(A=x,y,z;\, J=1,\dots, N-1) ,
\\
\hat{\cal K}\sigma_N^{A}\hat{\cal K}^{-1}
= \re^{+\ri\pi{\tt k}\sigma_1^z} \sigma^{A}_{1} \re^{-\ri\pi{\tt k}\sigma_1^z}.
\end{gather*}
In turn, the adjoint action of ${\cal K}$ on the transfer matrix~\eqref{Tmat1} results in
a cyclic permutation of the inhomogeneities:
\begin{gather}\label{Kact1a}
\hat{\cal K}
{\mathbb T}(\zeta\,|\,\eta_N,\eta_{N-1},\dots,\eta_1)\hat{\cal
 K}^{-1}=
{\mathbb T}(\zeta\,|\,\eta_{N-1},\eta_{N-2},\dots,\eta_N).
\end{gather}
This relation shows that with the conditions~\eqref{aoisd981221} and~\eqref{iisasausa} imposed,
the $r$-site translation operator belongs to the commuting family,
\begin{gather*}
{\mathbb K}=\hat{\cal K}^r\colon\quad
{\mathbb K}^L=\re^{2\pi\ri{\tt k}\mathbb{S}^z},\qquad
[\mathbb{K},\mathbb{T}(\zeta)]=0.
\end{gather*}
In fact $\mathbb{K}$ not only commutes, but it can be expressed in terms of the
transfer matrix. Assuming the normalization~\eqref{normT} for $\mathbb{T}(\zeta)$
and taking into account~\eqref{cond1}, it's straightforward to show
\begin{eqnarray}\label{Kdef1a}
{\mathbb K}=(q-q^{-1})^{-N} \bigg[\prod_{1<m<\ell<r}
\bigg(q^2+q^{-2}-\frac{\eta_\ell}{\eta_m}-\frac{\eta_m}{\eta_\ell}\bigg)^{-L}\bigg]
\prod_{\ell=1}^r q^{\frac{N}{2}} \mathbb{T}(-q^{-1}\eta_\ell).
\end{eqnarray}
It follows that the eigenvalue of the $r$-site translation operator on the Bethe state~\eqref{Bstate1}
 is given~by
\begin{gather*}
\mathbb{K}\bm{\Psi}=K\bm{\Psi}\colon\quad
K=\re^{\ri\pi r {\tt k}}q^{rS^z-\frac{rN}{2}}\prod_{\ell=1}^r
\frac{A_+(-q^{+1}\eta_\ell)}{A_+(-q^{-1}\eta_\ell)},
\end{gather*}
where $A_+(\zeta)$ is the same as in~\eqref{Qdef1A}.

\subsection{Quasi-shift operators}
In~\cite{Ikhlef:2011ay}, which was devoted to the study of the
${\cal Z}_2$ invariant model, the authors introduced the so-called quasi-shift operator.
It turns out to be a key player for the description of the scaling limit.
Similar operators may be defined for the general model provided the
restrictions~\eqref{aoisd981221} and~\eqref{iisasausa} are imposed.
They are likewise expected to be important
for analyzing the critical behaviour.

The construction of the quasi-shift operators is based on the following observation.
When the restrictions~\eqref{iisasausa} on the $\eta_J$ are imposed, the cyclic permutation of
the inhomogeneities within each group
$(\eta_{J+r},\eta_{J+r-1},\dots,\eta_{J+1})\mapsto (\eta_{J+1},\eta_{J+r},\dots,\eta_{J+2})$
with $J=\ell ({\rm mod}\, r)$ is equivalent to an overall cyclic permutation of the ordered set
$(\eta_N,\eta_{N-1},\dots, \eta_1)$. In view of equation~\eqref{YBE321}, this implies that
there exists $r$ inequivalent operators $\hat{\cal D}^{(\ell)}$, such that
\begin{gather}
\label{ZT-action}
\big(\hat{\cal D}^{(\ell)}\big)^{-1} {\mathbb T}(\zeta\,|\,\eta_N,\eta_{N-1},\dots,\eta_1)
\hat{\cal D}^{(\ell)}=
{\mathbb T}(\zeta\,|\,\eta_{N-1},\eta_{N-2},\dots,\eta_N),\qquad \ell=1,2,\dots,r.
\end{gather}
Explicitly, they are given by
\begin{gather}
\label{zelldef}
\hat{\cal D}^{(\ell)}=\prod_{j=0}^{L-1}\bigg(\hat{{\mathcal K}}^{\ell +rj}
\bigg[\overset{\displaystyle \curvearrowleft}{\prod_{m=1}^{r-1}}
\check{\bm{R}}_{m+1,m}\big(\eta_{\ell}/\eta_{\ell+m}\big)\bigg]
\hat{{\mathcal K}}^{-\ell -rj}\bigg),
\end{gather}
where the outer product is unordered since the factors inside it
commute with each other. It is worth mentioning that the operator
\begin{gather*}
\hat{\cal D}^{(r)}=\overset{\displaystyle \curvearrowleft}{\prod_{m=1}^{N-1}}
\check{\bm{R}}_{m+1,m}\big(\eta_{r}/\eta_{m}\big)
\end{gather*}
plays a special r\^ole.
Though the adjoint action of $\hat{\cal D}^{(\ell)}$ on the transfer matrix
results in~a~permutation of the inhomogeneities as in~\eqref{ZT-action}, a~similar
formula does not hold for the entries $\hat{\mathsf A}$,
$\hat{\mathsf B}$, $\hat{\mathsf C}$ and $\hat{\mathsf D}$ of the monodromy matrix.
However, it turns out that
\begin{gather}\label{ZM-action2a}
\big(\hat{\cal D}^{(r)}\big)^{-1}{\bm M}(\zeta\,|\,\eta_N,\eta_{N-1},\dots,\eta_1)
\hat{\cal D}^{(r)}={\bm M}(\zeta\,|\,\eta_{N-1},\eta_{N-2},\dots,\eta_N).
\end{gather}

Combining equations~\eqref{Kact1a} and~\eqref{ZT-action} it is easy to see that the operators
\begin{gather}\label{Bdef0}
 \mathbb{K}^{(\ell)}=\hat{\cal D}^{(\ell)} \hat{\cal K},\qquad \ell=1,2,\dots,r,
\end{gather}
belong to the commuting family,
\begin{gather*}
\big[\mathbb{K}^{(\ell)},\mathbb{T}(\zeta)\big]=0.
\end{gather*}
Moreover, their product coincides with the $r$-site translation:
\begin{gather*}
\prod_{\ell=1}^r\mathbb{K}^{(\ell)}=\mathbb{K}.
\end{gather*}
We will refer to $\mathbb{K}^{(\ell)}$ as the quasi-shift operators.\footnote{There is a slight difference between the terminology used here and that from
the work~\cite{Ikhlef:2011ay}. The latter focuses on
 the ${\cal Z}_2$ invariant model, where there are two operators $\mathbb{K}^{(1)}$ and $\mathbb{K}^{(2)}$
such that $\mathbb{K}^{(1)}\mathbb{K}^{(2)}=\mathbb{K}$. They introduce the
``quasi-momentum'', which essentially coincides with
the logarithm of $\mathbb{K}^{(1)}\big(\mathbb{K}^{(2)}\big)^{-1}$. The term ``quasi-shift''
was coined in~\cite{Frahm:2013cma}.\label{ft3}}
Similar as in equation~\eqref{Kdef1a}, it is possible to express them
in terms of the transfer matrix as
\begin{gather}\label{Bdef1a}
{\mathbb K}^{(\ell)}=q^{-\frac{N}{2}}
\bigg(\prod_{J=1}^N
\big(\eta_J-q^{-2}\eta_{\ell}\big)^{-1}\bigg)
\mathbb{T}(-q^{-1}\eta_{\ell}).
\end{gather}
Their eigenvalues for the Bethe state~\eqref{Bstate1} are given by
\begin{gather*}
{K}^{(\ell)}=\re^{\ri\pi{\tt k}}q^{S^z-N/2}
\frac{A_+\big(-q^{+1}\eta_{\ell}\big)}{A_+(-q^{-1}\eta_{\ell})}
\end{gather*}
with $A_+(\zeta)$ from~\eqref{Qdef1A}.

\subsection{Hamiltonians}
In the case of the homogeneous six-vertex model the transfer matrix
commutes with the spin $\frac{1}{2}$ Heisenberg $XXZ$ Hamiltonian.
A similar important property holds true for the
model, where the parameters $\eta_J$
satisfy the periodicity conditions~\eqref{iisasausa} with any $r\ge 1$.
Namely the corresponding commuting family
contains spin chain Hamiltonians
that are given by a sum of
terms, each of which is built out of local spin operators from $r+1$ consecutive sites of the lattice.
There are $r$ such Hamiltonians and, in terms of the row-to-row transfer matrix, they are expressed as
\begin{gather}\label{Helldef1}
\mathbb{H}^{(\ell)}=
2\ri\zeta\partial_\zeta\log\big(\mathbb{T}\big({-}q^{-1}\zeta\big)\big)\big|_{\zeta=\eta_\ell}
-2\ri \sum_{J=1}^N(1-q^2\eta_J/\eta_{\ell})^{-1}, \qquad \ell=1,2,\dots,r.
\end{gather}
A calculation shows that
\begin{gather*}
\mathbb{H}^{(\ell)}=\sum_{m=0}^{L-1} \mathbb{K}^{m} \hat{\mathsf{H}}^{(\ell)} \mathbb{K}^{-m},
\end{gather*}
where $\hat{\mathsf H}^{(\ell)}\in{\rm End}\big(\mathscr{V}_N\big)$ are built
from the $\sigma^A_J$ with $J=\ell,\ell+1,\dots,\ell+r$.
Assuming that $L\ge 2$ so that the sites $\ell$ and $\ell+r$ are not identified with one another,
 the operators $\hat{\mathsf H}^{(\ell)}$ are given by
\begin{gather*}
\hat{\mathsf{H}}^{(\ell)}=\sum_{m=1}^{r} \hat{\mathsf{S}}_m^{({\ell})}
\hat{\mathsf{J}}_{\ell+m,\ell+m-1}(\eta_{\ell}/\eta_{\ell+m-1})
\big(\hat{\mathsf{S}}^{({\ell})}_m\big)^{-1}.
\end{gather*}
Here we use the notation
\begin{gather*}
\hat{\!\mathsf{J}}_{n+1,n}(\zeta)=
2\ri\zeta \partial_\zeta\log(\check{\bm R}_{n+1,n}(\zeta))
\end{gather*}
and
\begin{gather*}
\hat{\mathsf{S}}^{(\ell)}_{m}=
\overset{\displaystyle \curvearrowleft}{\prod_{n=\ell+m}^{\ell+r-1}}
\check{\bm R}_{n+1,n}({\eta_\ell}/{\eta_{n}}),\qquad
 \hat{\mathsf S}^{(\ell)}_{r}=\hat{\bf 1},
\end{gather*}
with $\check{\bm R}_{n+1,n}$ as in~\eqref{Rcheckdef1}.

\subsection[Interplay with CP, T and Hermitian conjugation]{Interplay with ${\cal CP}$, ${\cal T}$ and Hermitian conjugation}
In order to incorporate the ${\cal C}$, ${\cal P}$ and ${\cal T}$ conjugations
for the model with translational invariance, extra constraints need to be imposed
on the parameters in addition to~\eqref{iisasausa}. For the three cases
considered in Section~\ref{sec4}, a brief examination shows that the
first and second ones~\eqref{case1a},~\eqref{case2a} are compatible with
the periodicity condition $\eta_{J+r}=\eta_J$ while the last case~\eqref{case3} is not.
Assuming the inhomogeneities are as in $(i)$ or $(ii)$,
the action of ${\cal CP}$ and ${\cal T}$ on the quasi-shift operators and the Hamiltonians
is given by
\begin{gather*}
\hat{\cal C}\hat{\cal P}\mathbb{H}^{(\ell)}\hat{\cal C}\hat{\cal P} =
\mathbb{H}^{(r-\ell+1)}, \qquad
\hat{\cal C}\hat{\cal P}{\mathbb K}^{(\ell)}\hat{\cal C}\hat{\cal P}
= \big({\mathbb K}^{(r-\ell+1)}\big)^{-1}
\end{gather*}
and
\begin{gather*}
\text{case $(i)$}\colon\qquad
\hat{\cal T}\mathbb{H}^{(\ell)}\hat{\cal T}=\mathbb{H}^{(\ell)},\qquad
\hat{\cal T}{\mathbb K}^{(\ell)}\hat{\cal T}= {\mathbb K}^{(\ell)},\nonumber
\\
\text{case $(ii)$}\colon\qquad
\hat{\cal T}\mathbb{H}^{(\ell)}\hat{\cal T}=\mathbb{H}^{(r-\ell+1)},\qquad
\hat{\cal T}{\mathbb K}^{(\ell)}\hat{\cal T}= {\mathbb K}^{(r-\ell+1)}.
\end{gather*}
These follow from formulae~\eqref{Bdef1a} and~\eqref{Helldef1}, which express
 $\mathbb{K}^{(\ell)}$ and $\mathbb{H}^{(\ell)}$ in terms of $\mathbb{T}(\zeta)$, and
equations~\eqref{CP},~\eqref{Ttrans} that describe the commutation relations
of $\hat{\cal C}\hat{\cal P}$ and $\hat{\cal T}$ with the transfer matrix.

The Hermitian conjugation~\eqref{star1} that is consistent with the integrable structure
involves the matrix $\hat{{\mathsf X}}$, which depends on the reality conditions
imposed on the set $\big\{\eta_J^*\big\}_{J=1}^N=\big\{\eta_J\big\}_{J=1}^N$.
In~the case $(i)$, the
expressions~\eqref{Xcase1},~\eqref{Sdef1b} can be simplified if the $r$-site periodicity conditions~\eqref{iisasausa} are taken into account. It turns out that
\begin{gather}\label{xrsite1a}
\hat{\mathsf{X}}=\prod_{m=0}^{L-1}\mathbb{K}^m \hat{\mathsf{X}}^{(1)} \mathbb{K}^{-m},\qquad
\hat{\mathsf{X}}^{(1)}=\bigg(\prod_{\ell=1}^r (\eta_\ell)^{\frac{1}{2}\sigma_\ell^z}\bigg)
\big(\hat{\mathsf{S}}_2^{({1})}\big)^{-1},
\end{gather}
where
\begin{gather*}
\hat{\mathsf{S}}_2^{({1})}=\overset{\displaystyle \curvearrowleft}{\prod_{m=2}^{r}}
\bigg[\,\overset{\displaystyle \curvearrowleft}{\prod_{n=r-m+1}^{r-1}}
\check{\bm{R}}_{n+1,n}(\eta_{m}/\eta_{n+m-r})\bigg].
\end{gather*}
Notice that each of the terms $\mathbb{K}^m \hat{\mathsf{X}}^{(1)} \mathbb{K}^{-m}$ entering into the product expression~\eqref{xrsite1a}
acts non-trivially only on sites $mr+1,mr+2,\dots, mr+r$.
For case $(ii)$ from Section~\ref{sec4}, $\hat{{\mathsf X}}$ is the
diagonal matrix given by equation~\eqref{oiasd9821}.

The formula for the $\star$-conjugation
also contains the operator $\mathbb{Y}$, subject to the condi\-ti\-ons~\eqref{Ycond1} and~\eqref{Ycond2}.
Independently of its choice $(\mathbb{T}(\zeta))^\star=\mathbb{T}(\zeta^*)$
from which it follows that
\begin{gather}
\text{case $(i)$}\colon\qquad
\big(\mathbb{H}^{(\ell)}\big)^\star=\mathbb{H}^{(r-\ell+1)},\qquad
\big({\mathbb K}^{(\ell)}\big)^\star= \big({\mathbb K}^{(r-\ell+1)}\big)^{-1}, \nonumber
 \\
\text{case $(ii)$}\colon\qquad
\big(\mathbb{H}^{(\ell)}\big)^\star=\mathbb{H}^{(\ell)},\qquad
\big({\mathbb K}^{(\ell)}\big)^\star= \big({\mathbb K}^{(\ell)}\big)^{-1}.
\label{aiusd98128912}
\end{gather}
Though the Hamiltonians and quasi-shift operators are not preserved under
the ${\cal C}{\cal P}$ transformation and the $\star$-conjugation,
the combinations
\begin{gather}
\label{aiosd901212}
\mathbb{H}=\sum_{\ell=1}^r\mathbb{H}^{(\ell)},\qquad
 \mathbb{K}=\prod_{\ell=1}^r\mathbb{K}^{(\ell)}
\end{gather}
satisfy the conditions
\begin{gather}
\label{asoid9812900912}
\hat{\cal C}\hat{\cal P}\mathbb{H}\hat{\cal C}\hat{\cal P}=\mathbb{H}, \qquad
\hat{\cal T}\mathbb{H}\hat{\cal T}=\mathbb{H},\qquad
\mathbb{H}^\star=\mathbb{H},\nonumber
 \\
\hat{\cal C}\hat{\cal P}{\mathbb K}\hat{\cal C}\hat{\cal P}={\mathbb K}^{-1},\qquad
\hat{\cal T}{\mathbb K}\hat{\cal T}= {\mathbb K}, \qquad
\mathbb{K}^\star=\mathbb{K}^{-1}
\end{gather}
for both cases $(i)$ and $(ii)$.

\section[Zr-invariance]
{${\cal Z}_{\boldsymbol r}$-invariance}\label{sec7}

The operators $\hat{\cal D}^{(\ell)}\in{\rm End}\big(\mathscr{V}_N\big)$,
acting on the transfer matrix,
result in a cyclic shift of the inhomogeneities as in equation~\eqref{ZT-action}.
Taking into account that $\eta_{J+r}=\eta_J$, after $r$ consecutive
applications of this transformation
the
inhomogeneities will return to their original order.
Thus~$\big(\hat{\cal D}^{(\ell)}\big)^r$ commutes with $\mathbb{T}(\zeta)$.
It turns out that if the inhomogeneities are specified to be
\begin{gather}\label{zsym1}
\eta_{J}=(-1)^r\re^{\frac{\ri\pi}{r} (2J-1)},
\end{gather}
then further
\begin{gather*}
\big(\hat{\cal D}^{(\ell)}\big)^r=\hat{\bf 1},\qquad
\ell=1,2,\dots,r.
\end{gather*}
In this case all the operators $\hat{\cal D}^{(\ell)}$ can be expressed in terms of
\begin{gather}
\label{aiousd890129012}
\hat{\cal D}\equiv \hat{\cal D}^{(r)}=\overset{\displaystyle \curvearrowleft}{\prod_{m=1}^{N-1}}
\check{\bm{R}}_{m+1,m}\big(\re^{-2\ri \pi m/r}\big)
\end{gather}
and the one-site translation operator ${\cal K}$~\eqref{aiosd901221}. Namely,
\begin{gather*}
\hat{\cal D}^{(\ell)}=\hat{{\cal K}}^{\ell} \hat{\cal D} \hat{\cal K}^{-\ell}
\end{gather*}
with
\begin{gather*}
\hat{\cal K}^r=\mathbb{K},\qquad
\hat{\cal D}^r=1.
\end{gather*}

When the condition~\eqref{zsym1} is imposed, both operators $\hat{\cal D}$
and $\hat{{\cal K}}$ preserve the commuting family. Their adjoint
action on the transfer matrix and $\mathbb{A}_\pm(\zeta)$
is given by the similar formulae
\begin{gather}
\hat{\cal D}^{-1}{\mathbb T}(\zeta)\hat{\cal D}=
{\mathbb T}\big(\re^{+2\pi\ri/r}\zeta\big),\qquad
\hat{\cal D}^{-1}{\mathbb A}_\pm(\zeta) \hat{\cal D}=
{\mathbb A}_\pm\big(\re^{+2\pi\ri/r}\zeta\big) ,\nonumber
\\
\hat{\cal K}^{-1}{\mathbb T}(\zeta) \hat{\cal K}=
{\mathbb T}\big(\re^{-2\pi\ri/r}\zeta\big),\qquad
\hat{\cal K}^{-1}{\mathbb A}_\pm(\zeta) \hat{{\cal K}}=
{\mathbb A}_\pm\big(\re^{-2\pi\ri/r}\zeta\big).
\label{poasd0-12A}
\end{gather}
Among others, these imply that
\begin{gather}\label{poasd0-12B}
\big[\hat{\cal D},{\mathbb S}^z\big]=\big[\hat{\cal D},{\mathbb H}\big]=
\big[\hat{\cal D},{\mathbb K}\big]=0,\qquad
\big[\hat{\cal K},{\mathbb S}^z\big]=\big[\hat{{\cal K}},{\mathbb H}\big]=
\big[\hat{\cal K},{\mathbb K}\big]=0
\end{gather}
and
\begin{gather}
\hat{\cal D}^{-1}\mathbb{H}^{(\ell)}\hat{\cal D}=\mathbb{H}^{(\ell+1)},\qquad
\hat{\cal D}^{-1}\mathbb{K}^{(\ell)}\hat{\cal D}=\mathbb{K}^{(\ell+1)},\qquad
\ell=1,2,\dots,r-1, \nonumber\\
\hat{\cal D}^{-1}\mathbb{H}^{(r)}\hat{\cal D}=\mathbb{H}^{(1)},\qquad
\hat{\cal D}^{-1}\mathbb{K}^{(r)}\hat{\cal D}=\mathbb{K}^{(1)}.\label{asido120129021}
\end{gather}
With $\mathbb{H}$~\eqref{aiosd901212} taken as the Hamiltonian, the system
possesses ${\cal Z}_r$ invariance (in the usual quantum mechanical sense) generated by
the operator $\hat{\cal D}\colon \hat{\cal D}^r=1$.
It deserves to be mentioned that in this case equation~\eqref{ZM-action2a} becomes
\begin{gather*}
\hat{\cal D}^{-1}\bm{M}(\zeta)\hat{\cal D}=\bm{M}\big(\re^{+2\pi\ri/r}\zeta\big).
\end{gather*}
With the latter at hand and
taking into account that $\hat{\cal D}\,|\,0\,\rangle=|\,0\,\rangle$,
it is straightforward to check that the
action of $\hat{\cal D}$ on the Bethe state~\eqref{Bstate1} is given by
\begin{gather*}
\hat{\cal D}\boldsymbol{\Psi}(\{\zeta_j\})=\boldsymbol{\Psi}\big(\{\re^{+2\pi\ri/r}\zeta_j\}\big).
\end{gather*}
Note that if $\{\zeta_m\}$ solves the Bethe ansatz equations~\eqref{bae} with
$\eta_J=(-1)^r\re^{\frac{\ri\pi}{r} (2J-1)}$ then the set $\{\re^{+2\pi\ri/r}\zeta_m\}$
is also a solution of the same equations.

To summarize, when the $\{\eta_J\}$ are fixed to be as in~\eqref{zsym1}
and with $|q|=|\omega|=1$, the inhomogeneous six-vertex model, together with
the ${\rm U}(1)$ symmetry generated by the operator $\mathbb{S}^z$, possesses a
set of discrete global symmetries whose generators are
$\hat{\cal D}$, $\hat{{\cal K}}$, $\hat{\cal C}\hat{\cal P}$ and
$\hat{\cal T}$. The latter obey the commutation relations
\begin{gather*}
\hat{\cal C}\hat{\cal P}\hat{\cal D}\hat{\cal C}\hat{\cal P}=\hat{\cal D}^{-1},\qquad
\hat{\cal T}\hat{\cal D}\hat{\cal T}=\hat{\cal D},
\\
\hat{\cal C}\hat{\cal P}\hat{\cal K}\hat{\cal C}\hat{\cal P}=\hat{\cal K}^{-1},\qquad
\hat{\cal T}\hat{\cal K}\hat{\cal T}=\hat{\cal K}.
\end{gather*}


\section{Examples}
Having discussed some general aspects of the lattice system, we now turn to the
two simplest cases -- the homogeneous and ${\cal Z}_2$ invariant models. The accent will be placed
on the properties specific to these models.

\subsection{Homogeneous six-vertex model}
In the homogeneous case all the parameters $\{\eta_J\}$ are equal and can be set to one:
\begin{gather*}
\eta_J=1, \qquad J=1,\dots,N.
\end{gather*}
Then it turns out that the operators $\mathbb{A}_\pm(\zeta)$ with $\zeta$ real are Hermitian
 w.r.t.~the standard Hermitian matrix conjugation:
\begin{gather*}
\big[\mathbb{A}_\pm(\zeta)\big]^\dag=\mathbb{A}_\pm(\zeta^*).
\end{gather*}
The latter follows from the general relation~\eqref{ddagBCTb} and the fact that
 the matrix $\hat{\mathsf X}$~\eqref{xrsite1a}, entering into
the conjugation condition~\eqref{ddag1}, becomes the identity
\begin{gather*}
 \hat{\mathsf X}=\hat{\bf 1}.
\end{gather*}
Since the sesquilinear form corresponding to the $\dagger$-conjugation is positive definite,
the eigenvalues of $\mathbb{A}_\pm(\zeta)$ are polynomials in $\zeta$ with real coefficients.
In consequence the set of zeroes of~$A_+(\zeta)$,
\begin{gather*}
A_+(\zeta)=\prod_{m=1}^M(1-\zeta/\zeta_m),
\end{gather*}
which solves the Bethe ansatz equations, coincides with the complex conjugated set:
\begin{gather*}
\{\zeta_m\}_{m=1}^M=\{\zeta^*_m\}_{m=1}^M.
\end{gather*}
In view of the relation~\eqref{CPTBethe1},
for the homogeneous model any Bethe state is invariant under the ${\cal CPT}$ conjugation
\begin{gather*}
\hat{\cal C}\hat{\cal P}\hat{\cal T}\bm{\Psi}=\bm{\Psi}.
\end{gather*}

The normalization of the Bethe state $\bm{\Psi}$~\eqref{Bstate1}
is different to that from~\cite{Gaudin:1981cyg}.
In the latter, the authors consider
\begin{eqnarray}\label{aoisd98129812}
\boldsymbol{\Psi}'=
\sum_{1\leq x_1<x_2<\cdots<x_M\leq N}\Psi'(x_1,\dots,x_M)\sigma_{x_M}^-\cdots
\sigma_{x_1}^- \,|\,0\,\rangle
\end{eqnarray}
with the wavefunction being
 \begin{gather*}
\Psi'(x_1,\dots,x_M)=\sum_{\hat P}A_{\hat P} \re^{\ri \sum_{m=1}^M p_{{\hat P}m}x_m}.
\end{gather*}
Here $A_{\hat P}$ is the same as in~\eqref{ap} and
\begin{gather*}
\re^{\ri p_m}=\frac{1+q\zeta_m}{q+\zeta_m}.
\end{gather*}
Notice that the function $\phi_m(x)$~\eqref{single} with $\eta_J=1$ may be written as
\begin{gather*}
\phi_m(x)=-\ri q^{\frac{1}{2}}\omega^{-1} \frac{\big(q-q^{-1}\big)\zeta_m}{1+q\zeta_m} \re^{\ri p_m x}.
\end{gather*}
This way the relation between the Bethe state defined through
\eqref{cba}--\eqref{single} and $\bm{\Psi}'$ reads
\begin{gather*}
\bm{\Psi}'=\alpha(\zeta_1,\dots,\zeta_M)\bm{\Psi}(\{\zeta_j\}),
\end{gather*}
where
\begin{gather*}
\alpha(\zeta_1,\dots,\zeta_M)=
\big({-}\ri q^{-\frac{1}{2}}\re^{-\ri\pi{\tt k}} \big(q-q^{-1}\big)\big)^{-M} A_+\big({-}q^{-1}\big).
\end{gather*}

The norm of the Bethe state $\bm{\Psi}'$~\eqref{aoisd98129812} w.r.t.~the positive definite inner product is given by
\begin{gather*}
\parallel\!\bm{\Psi}'\!\parallel^2=\sum_{1\leq x_1<x_2<\cdots<x_M\leq N}
\big|\Psi'(x_1,\dots,x_M)\big|^2.
\end{gather*}
The remarkable formula for this norm, originally conjectured by Gaudin, McCoy and Wu in~\cite{Gaudin:1981cyg} and proven in the work of Korepin~\cite{Korepin:1982gg}
can be written as
\begin{gather*}
\parallel\!\bm{\Psi}'\!\parallel^2=
\big|\alpha(\zeta_1,\dots,\zeta_M)\big|^{2}\mathfrak{K}[{\boldsymbol \Psi}]\prod_{m=1}^M\zeta_m,
\end{gather*}
where we use the notation $\mathfrak{K}[{\boldsymbol \Psi}]$ from~\eqref{FinalNorm}.
Notice that the r.h.s.\ of the relation coincides with~$\mathfrak{N}_Y[{\boldsymbol \Psi}]$~\eqref{KYnorm1}, where the operator $\mathbb{Y}$
is taken to be
\begin{gather}
\label{YYYYY72837kjdsa}
{\mathbb Y}=\mathbb{Y}^\dag= \big(\ri \big(q-q^{-1}\big)\big)^{2{\mathbb S}^z-N}
{\mathbb A}_+(-q^{+1}){\mathbb A}_+\big({-}q^{-1}\big).
\end{gather}
At the same time $\parallel\!\bm{\Psi}'\!\parallel^2 \equiv\big(\bm{\Psi}',\bm{\Psi}'\big)_\star$ for
the $\star$-conjugation given by~\eqref{star1} with $\mathbb{Y}=\hat{\bf 1}$.
This way,
\begin{gather*}
\mathfrak{N}_1[{\boldsymbol \Psi}']=\mathfrak{N}_Y[{\boldsymbol \Psi}].
\end{gather*}
The last formula illustrates a point previously touched on in Section~\ref{sec5}
(see the paragraph below equation~\eqref{uassysay}). Namely, that
 different operators $\mathbb{Y}$ entering into the definition
of the $\star$-conjugation
may lead to the same Hermitian structure.
Note that the local spin operators are Hermitian w.r.t.~the $\dag$-conjugation,
\begin{gather*}
\big(\sigma_J^A\big)^\dag=\sigma_J^A, \qquad
 A=x,y,z,\quad J=1,\dots,N.
\end{gather*}
For the $\star$-conjugation
\begin{gather*}
\big(\sigma^A_J\big)^\star=\mathbb{Y}^{-1} \sigma^A_J \mathbb{Y}
\end{gather*}
with $\mathbb{Y}$ as in equation~\eqref{YYYYY72837kjdsa},
when written explicitly in terms of the local spin operators,
would result in a highly cumbersome expression.

Finally in the case of the homogeneous model, which corresponds to $r=1$ in~\eqref{iisasausa},
there is one quasi-shift operator that coincides with the one-site
lattice translation:
\begin{gather*}
\mathbb{K}^{(1)}=\mathbb{K}=\hat{\cal K}.
\end{gather*}
The Hamiltonian $\mathbb{H}^{(1)}$~\eqref{Helldef1} is just the $XXZ$ spin chain Hamiltonian
\begin{gather*}
\mathbb{H}^{(1)}=-\frac{\ri}{q-q^{-1}}
\sum_{m=1}^N\big(\sigma_m^x\sigma_{m+1}^x+\sigma_m^y\sigma_{m+1}^y+\Delta
\big(\sigma_m^z\sigma_{m+1}^z-\hat{\bm{1}}\big)\big),\qquad
\Delta=\frac{1}{2}\big(q+q^{-1}\big)
\end{gather*}
subject to the quasi-periodic boundary conditions
\begin{gather*}
\sigma^x_{N+1}\pm\ri\sigma^y_{N+1}
=\re^{\pm2\pi\ri{\tt k}}\big(\sigma^x_{1}\pm\ri\sigma^y_{1}\big),\qquad
\sigma_{N+1}^z=\sigma^z_1.
\end{gather*}

\subsection[Z2 invariant model]{${\cal Z}_{\boldsymbol2}$ invariant model}
The model corresponds to the case when the number of lattice columns $N$ is even
and the inhomogeneities are given by equation~\eqref{zsym1} with $r=2$:
\begin{gather*}
\eta_J=\ri(-1)^{J-1}.
\end{gather*}

{\bf Bethe ansatz equations.}
For the ${\cal Z}_2$ invariant model, the algebraic system~\eqref{bae} redu\-ces~to
\begin{gather*}
\bigg(\frac{1+q^{+2}\zeta^2_m}{1+q^{-2}\zeta^2_m }\bigg)^{\frac{N}{2}}
=-\omega^2q^{2S^z}\prod_{j=1}^M\frac{\zeta_j-q^{+2}\zeta_m}{\zeta_j-q^{-2}\zeta_m},\qquad
m=1,2,\dots,M,
\end{gather*}
where $M=N/2-S^z$. The eigenvalue of the transfer matrix~\eqref{tmat} for the Bethe
state~\eqref{Bstate1} is given by\footnote{The transfer matrix ${\mathbb T}(\beta)$ and the Hamiltonian
defined in equations~(5) and~(2)
of~\cite{Bazhanov:2019xvy}, respectively, coincide with
$\big(q+q^{-1} \zeta^2\big)^{-N/2}\hat{\mathsf V}{\mathbb T}(\zeta)\hat{\mathsf V}^{-1}$,
$\hat{\mathsf{V}}\mathbb{H}\hat{\mathsf{V}}^{-1}$,
where $\mathbb{T}(\zeta)$ is given by~\eqref{tmat},
$\mathbb{H}$ is as in~\eqref{halt} below,
the matrix
$\hat{{\mathsf V}}=\prod_{m=1}^N\exp\big(\frac{\ri\pi}{4}\sigma^z_{2m-1}\big)$,
while $\zeta=\re^{-2\beta}$.}
\begin{gather*}
T(\zeta)=\omega^{+1}q^{+S^z}\big(1+q^{-2}\zeta^2\big)^{\frac{N}{2}}
\prod_{m=1}^M\frac{\zeta_m-q^{+2}\zeta}{\zeta_m-\zeta}
+\omega^{-1}q^{-S^z}\big(1+q^{+2}\zeta^2\big)^{\frac{N}{2}}
\prod_{m=1}^M\frac{\zeta_m-q^{-2}\zeta}{\zeta_m-\zeta}.
\end{gather*}
The parameters $q$ and $\omega$ are assumed to be
unimodular, i.e.,
\begin{gather*}
q=\re^{\ri\gamma},\qquad 0<\gamma<\pi;\qquad
\omega=\re^{\ri\pi{\tt k}},\qquad -\frac12<{\tt k}\le \frac12.
\end{gather*}
In~\cite{Bazhanov:2019xvy,Bazhanov:2020}
the parameter $\gamma$ is swapped for $n$, such that
\begin{gather*}
\gamma=\frac{\pi}{n+2}.
\end{gather*}

{\bf Hamiltonians.} Let's denote by ${\mathbb H}^{(+)}={\mathbb H^{(1)}}$ and ${\mathbb
 H}^{(-)}={\mathbb H^{(2)}}$ the Hamiltonians
defined by equation~\eqref{Helldef1} with~$r=2$.
It is straightforward to derive the following expressions:
\begin{gather*}
{\mathbb H}^{(+)}=\frac{1}{2}\tan(\gamma)
\sum_{m=1}^{N}\sigma^z_m\sigma^z_{m+1}-\frac{1}{\sin(2\gamma)}\sum_{m=1}^{N/2}
\big(\sigma^x_{2m}\sigma^x_{2m+2}+\sigma^y_{2m}\sigma^y_{2m+2}+
\sigma^z_{2m}\sigma^z_{2m+2}\big)
\\ \hphantom{{\mathbb H}^{(+)}=}
{}+\frac{\ri}{2\cos(\gamma)}\!\sum_{m=1}^{N/2}\big(\big(\sigma_{2m}^x\sigma_{2m+1}^x\!+\!
\sigma_{2m}^y\sigma_{2m+1}^y\big)\sigma^z_{2m+2} \!-\!\sigma^z_{2m}
\big(\sigma_{2m+1}^x\sigma_{2m+2}^x\!+\!\sigma_{2m+1}^y\sigma_{2m+2}^y\big)\big)
\\ \hphantom{{\mathbb H}^{(+)}=}
{}+\frac{N}{2}\cot(2\gamma) \hat{{\bf 1}}
\end{gather*}
and
\begin{gather*}
{\mathbb H}^{(-)}=\frac{1}{2}\tan(\gamma)\sum_{m=1}^{N} \sigma^z_m\sigma^z_{m+1}-
\frac{1}{\sin(2\gamma)}\sum_{m=1}^{N/2}\big(\sigma^x_{2m-1}\sigma^x_{2m+1}
+\sigma^y_{2m-1}\sigma^y_{2m+1}+\sigma^z_{2m-1}\sigma^z_{2m+1}\big)
\\ \hphantom{{\mathbb H}^{(-)}=}
{}+\frac{\ri}{2\cos(\gamma)}\sum_{m=1}^{N/2}\big(\big(\sigma_{2m-1}^x\sigma_{2m}^x\!+\!
\sigma_{2m-1}^y\sigma_{2m}^y\big)\sigma^z_{2m+1} \!-\!
\sigma^z_{2m-1}\big(\sigma_{2m}^x\sigma_{2m+1}^x\!+\!
\sigma_{2m}^y\sigma_{2m+1}^y\big)\big)
\\ \hphantom{{\mathbb H}^{(-)}=}
{}+\frac{N}{2}\cot(2\gamma)\hat{\bf 1},
\end{gather*}
where
\begin{gather*}
\sigma^x_{N+\ell} \pm \ri \sigma^y_{N+\ell}=\re^{\pm2\pi\ri{\tt k}}
\big(\sigma^x_{\ell} \pm \ri \sigma^y_{\ell}\big), \qquad
\sigma_{N+\ell}^z=\sigma^z_\ell, \qquad
\ell=1,2.
\end{gather*}
The eigenvalues of these operators for the Bethe state~\eqref{Bstate1}
are expressed through the set~$\{\zeta_m\}\!$~as
\begin{gather*}
{\mathbb H}^{(\pm)}\boldsymbol{\Psi}={\cal E}^{(\pm)}\boldsymbol{\Psi}\colon\quad
{\cal E}^{(\pm)}=\pm\!\!\sum_{m=1}^{N/2-S^z}
\frac{2(q-q^{-1})}{\zeta_m-\zeta^{-1}_m\mp\ri(q+q^{-1})}.
\end{gather*}
The Hamiltonian ${\mathbb H}={\mathbb H}^{(+)}
+{\mathbb H}^{(-)}$~\eqref{aiosd901212}, in terms of the local spin operators, takes the form
\begin{gather}
{\mathbb H}=\frac{1}{\sin(2\gamma)}\sum_{m=1}^{N}
\big(2\sin^2(\gamma) \sigma^z_m \sigma^z_{m+1}-
\big(\sigma^x_m \sigma^x_{m+2}+\sigma^y_m \sigma^y_{m+2}+
\sigma^z_m \sigma^z_{m+2}\big)\nonumber
\\ \hphantom{{\mathbb H}=}
{}+\ri \sin(\gamma)\big(\sigma_m^x\sigma_{m+1}^x+
\sigma_m^y\sigma_{m+1}^y\big)\big(\sigma^z_{m+2}-\sigma^z_{m-1}\big) \big)
+N\cot(2\gamma) \hat{{\bf 1}}.
\label{halt}
\end{gather}
Notice that $\mathbb{H}^{(\pm)}$ and $\mathbb{H}$
are given by a sum of terms,
where the local spin operators couple with their
nearest and next-to-nearest neighbours only.

\begin{figure}\centering
\vspace{0.2cm}
\scalebox{0.85}{
\includegraphics[width=11cm]{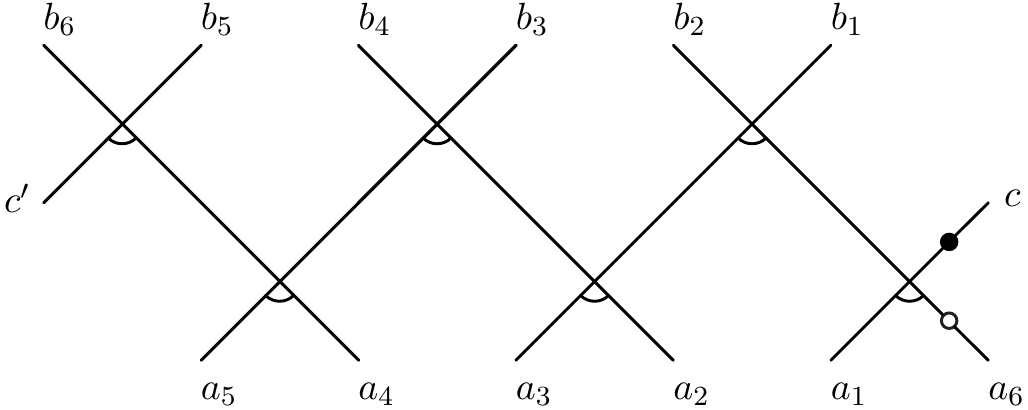}}
\caption{A graphical representation of the
matrix elements
$({\mathbb B})_{a_{N}a_{N-1}\dots a_1}^{b_{N}b_{N-1}\dots b_1}$
of the quasi-shift operator~\eqref{aospid91p021} for the chain of length $N=6$.
Summation over the spin indices assigned to internal edges is assumed.
The black and white dots correspond to $\omega^{c}$ and $\omega^{-a_6}$, respectively. Note that, since quasi-periodic
boundary conditions are imposed, $c$ and $c'$ should be identified.}
\label{pic3}
\end{figure}

{\bf Quasi-shift operators.}
For the ${\cal Z}_2$ invariant model there are two quasi-shift operators
$\mathbb{K}^{(+)}\equiv\mathbb{K}^{(1)}$ and
$\mathbb{K}^{(-)}\equiv\mathbb{K}^{(2)}$ \eqref{Bdef1a},
whose product is equal to the two-site translation operator
\begin{gather*}
{\mathbb K}=\mathbb{K}^{(+)} \mathbb{K}^{(-)}.
\end{gather*}
In this case it is convenient to define (see also footnote \ref{ft3})
\begin{gather*}
\mathbb{B}=\mathbb{K}^{(+)}\big(\mathbb{K}^{(-)}\big)^{-1}.
\end{gather*}
A computation based on equations~\eqref{zelldef} and~\eqref{Bdef0} shows that
\begin{gather}
\mathbb{B}=\omega^{-\sigma^z_{N}} \check{\bm R}_{N-1,N-2}(-1)
\check{\bm R}_{N-3,N-4}(-1)\cdots\check{\bm R}_{1,N}(-1)\nonumber
\\ \hphantom{\mathbb{B}=}
{}\times \omega^{+\sigma^z_{N}} \check{\bm R}_{N,N-1}(-1)
\check{\bm R}_{N-2,N-3}(-1)\cdots\check{\bm R}_{2,1}(-1).
\label{aospid91p021}
\end{gather}
This implies that $\mathbb{B}$
coincides with the transfer matrix of a homogeneous
six-vertex model on the $45^{\circ}$-rotated square
lattice with quasi-periodic boundary conditions (see Fig.~\ref{pic3}).
Its eigenvalues are given by
\begin{gather*}
{\mathbb B} \boldsymbol{\Psi}=B\boldsymbol{\Psi}\colon\quad
{B}=\prod_{m=1}^{N/2-S^z}\frac{(\zeta_m+\ri q)
\big(\zeta_m-\ri q^{-1}\big)}{(\zeta_m-\ri q)\big(\zeta_m+\ri q^{-1}\big)}.
\end{gather*}

{\bf Discrete symmetries.}
The generators of the ${\cal C}$, ${\cal P}$ and ${\cal T}$ conjugations
are given by the formulae~\eqref{Cconj1},~\eqref{oaisodi899832} and~\eqref{T1}, respectively,
with $\eta_J=\ri(-1)^{J-1}$.
Their adjoint action on the Hamiltonians $\mathbb{H}^{(\pm)}$
and quasi-shift operators $\mathbb{K}^{(\pm)}$ reads as
\begin{gather*}
\hat{\cal C}\hat{\cal P} \mathbb{H}^{(\pm)} \hat{\cal C}\hat{\cal P}=
\mathbb{H}^{(\mp)},\qquad
\hat{\cal T} \mathbb{H}^{(\pm)} \hat{\cal T}=\mathbb{H}^{(\pm)},
\\
\hat{\cal C}\hat{\cal P} \mathbb{K}^{(\pm)} \hat{\cal C}\hat{\cal P}=
\mathbb{K}^{(\mp)},\qquad
\hat{\cal T} \mathbb{K}^{(\pm)} \hat{\cal T}=\mathbb{K}^{(\pm)},
\end{gather*}
while the similar relations for $\mathbb{H}$ and $\mathbb{K}$
were already quoted in equation~\eqref{asoid9812900912}. Also it is easy to see
that
\begin{gather*}
\hat{\cal C}\hat{\cal P} \mathbb{B} \hat{\cal C}\hat{\cal P}=
\hat{\cal T} \mathbb{B} \hat{\cal T}=\mathbb{B}.
\end{gather*}

The lattice system possesses an additional ${\cal Z}_2$ symmetry,
whose generator is defined by equation~\eqref{aiousd890129012} with~$r=2$, i.e.,
\begin{gather}\label{Dz2case}
\hat{\cal D}= \prod_{m=1}^{N/2} \check{\bm R}_{2m,2m-1}(-1)\colon\quad \hat{\cal D}^2=1.
\end{gather}
One can show that the adjoint action of $\hat{\cal D}$ on the local spin operators is given by
\begin{gather*}
\hat{\cal D}\sigma^\pm_m\hat{\cal D}=\frac{1}{\cos(\gamma)}
\big(\sigma^{\pm}_{m+1}-\ri \sin(\gamma)\sigma^z_{m+1}\sigma^{\pm}_m\big),
\\
\hat{\cal D}\sigma^z_m\hat{\cal D}=\frac{1}{\cos^2(\gamma)}
\big(\sigma^z_{m+1}-\sin^2(\gamma)\sigma^z_m+2\ri \sin(\gamma)
\big(\sigma^+_{m+1}\sigma^-_m+\sigma^-_{m+1}\sigma^+_m\big)\big)
\end{gather*}
for odd values of $m$ and
\begin{gather*}
\hat{\cal D}\sigma^\pm_m\hat{\cal D}
=\frac{1}{\cos(\gamma)} \big(\sigma^{\pm}_{m-1}+\ri \sin(\gamma)
\sigma^{\pm}_m\sigma^{z}_{m-1}\big),
\\
\hat{\cal D}\sigma^z_m\hat{\cal D}=\frac{1}{\cos^2(\gamma)}
\big(\sigma^z_{m-1}-\sin^2(\gamma)\sigma^z_m-2\ri\sin(\gamma)
\big(\sigma^+_{m}\sigma^-_{m-1}+\sigma^-_{m}\sigma^+_{m-1}\big)\big)
\end{gather*}
for even $m$, while
\begin{gather*}
{\cal \hat D}\,|\,0\,\rangle=|\,0\,\rangle.
\end{gather*}
The commutation relations of ${\cal \hat D}$ with the operators from the commuting family
are provided by equations~\eqref{poasd0-12A} with $\re^{2\pi\ri/r}=-1$ and~\eqref{poasd0-12B}.
As for the Hamiltonians ${\mathbb H}^{(\pm)}$ and quasi-shift operators ${\mathbb K}^{(\pm)}$,
formula~\eqref{asido120129021} implies
\begin{gather*}
\hat{\cal D}{\mathbb H}^{(\pm)}\hat{\cal D}={\mathbb H}^{(\mp)},\qquad
\hat{\cal D}{\mathbb K}^{(\pm)}\hat{\cal D}={\mathbb K}^{(\mp)},\qquad
{\rm whereas} \qquad
\hat{\cal D}{\mathbb B}\hat{\cal D}=\mathbb{B}^{-1}.
\end{gather*}

{\bf Hermitian structures.}
Consider the Hermitian matrix $\hat{{\mathsf X}}$ entering into the conjugation con\-di\-tion~\eqref{ddag1}.
It is given by equation~\eqref{xrsite1a}, which for $\eta_J=\ri(-1)^{J-1}$ simplifies to
\begin{gather*}
\hat{\mathsf X}=
\bigg(\prod_{J=1}^N (\eta_J)^{\frac{1}{2}\sigma^z_J}\bigg)
\hat{\cal D}=\Sigma^z\hat{\cal D}\re^{\frac{\ri\pi}{2}({\mathbb S}^z-N/2)}.
\end{gather*}
Here $\Sigma^z$ stands for the diagonal matrix
\begin{gather*}
\Sigma^z=\sigma_N^z\otimes\sigma_{N-2}^z\otimes\cdots\otimes\sigma_2^z.
\end{gather*}
Note that the Hermiticity of $\hat{{\mathsf X}}$ follows from the relation
\begin{gather*}
\Sigma^z {\cal\hat D}^{\dag} \Sigma^z=
\re^{{\ri\pi}({\mathbb S}^z-N/2)} {\cal\hat D},
\end{gather*}
which can be easily verified using~\eqref{Dz2case}.

Unlike the homogeneous case, the matrix $\hat{{\mathsf X}}$ is non-trivial.
Among others, this implies that the operators ${\mathbb T}(\zeta)$
and ${\mathbb A}_\pm(\zeta)$ are not Hermitian w.r.t.~the standard matrix Hermitian conjugation.
Of the Hermitian structures consistent with the integrable structure, a special
r\^{o}le is played by~the one defined through the $\star$-conjugation~\eqref{star1} with
\begin{gather}
\label{ystar}
{\mathbb Y}=\re^{{\ri\pi}({\mathbb S}^z-N/2)} {\mathbb A}_+^{(\infty)}.
\end{gather}
As was already mentioned, this matrix satisfies the conditions~\eqref{Ycond1} and~\eqref{Ycond2}. The former can also be independently checked using the relations
\begin{gather*}
\Sigma^z \big({\mathbb A}_+^{(\infty)}\big)^\dag \Sigma^z=
{\cal\hat D} \big({\mathbb A}_+^{(\infty)}\big)^\dag{\cal\hat D}
=\re^{{\ri\pi}({\mathbb S}^z-N/2)} {\mathbb A}_+^{(\infty)}.
\end{gather*}
Recall that the ``norm'' of the Bethe state associated
with the $\star$-conjugation for the case~\eqref{ystar}
is given by
\begin{gather*}
\big({\cal \hat C \hat P \hat T} {\boldsymbol \Psi},{\boldsymbol \Psi}\big)_\star
={\mathfrak K}[{\bm \Psi}]\qquad
\big({\mathbb Y}=\re^{{\ri\pi}({\mathbb S}^z-N/2)}{\mathbb A}_+^{(\infty)}\big)
\end{gather*}
with ${\mathfrak K}[{\bm \Psi}]$ being defined by equation~\eqref{FinalNorm}.

The specialization of~\eqref{aiusd98128912}
to the ${\cal Z}_2$ invariant model results in
\begin{gather*}
\big({\mathbb H}^{(\pm)}\big)^\star={\mathbb H}^{(\mp)},\qquad
\big({\mathbb K}^{(\pm)}\big)^\star=\big({\mathbb K}^{(\mp)}\big)^{-1}
\end{gather*}
and therefore
\begin{gather*}
{\mathbb H}^\star={\mathbb H},\qquad
{\mathbb K}^\star={\mathbb K}^{-1},\qquad
{\mathbb B}^\star={\mathbb B}.
\end{gather*}

\section{Conclusion}
The subject matter of this paper is the integrable inhomogeneous six-vertex model on the square lattice.
The discussion is focused on various algebraic properties of the model, which
are important for studying the scaling limit.

We summarized the results concerning the diagonalization problem for the transfer matrix.
Explicit formulae for the Baxter $Q$-operators were presented as
special transfer matrices associated with the infinite dimensional
representations of the $q$-oscillator algebra.

The ${\cal C}$, ${\cal P}$
and ${\cal T}$ conjugations were introduced, and we described the constraints
on the parameters of the model such that these are consistent
with the integrable structure.
The~spe\-cial features of the model possessing translational
invariance were discussed. Among these, it~was pointed out that the commuting family contains
a set of quasi-shift operators $\mathbb{K}^{(\ell)}$ as~well as Hamiltonians
$\mathbb{H}^{({\ell})}$. The latter are distinguished in that they are
 given by a sum of~``local'' operators.
We also formulated the conditions for which the translationally invariant model
possesses an extra ${\cal Z}_r$ global symmetry and described its basic properties.

Both $\mathbb{H}^{({\ell})}$ and $\mathbb{K}^{(\ell)}$
are expected to play a key r\^ole for the study of the critical behaviour.
The~Hamiltonians are sparse matrices
(most of their elements are vanishing) and
there exist efficient algorithms for
finding the eigenvectors and eigenvalues belonging to the low energy part of the spectrum.
For a given eigenvector, it is straightforward to compute
the eigenvalue of the $Q$-operators and, in turn,
obtain the corresponding solution of the Bethe ansatz equations.
In~our subsequent paper~\cite{Bazhanov:2020} this approach is used for analyzing the scaling limit
of the homogeneous and ${\cal Z}_2$ invariant six-vertex models.
The explicit formulae collected in the last section form the starting point of that work.

For the case when both the anisotropy and twist parameters are unimodular,
$|q|=|\omega|=1$, as well as when the set of inhomogeneities $\{\eta_J\}$ coincides with the complex
conjugated set~$\{\eta_J^*\}$,
a family of Hermitian structures were introduced, which are consistent
with the integrable one. Again, this analysis forms the preliminary set up
for the study of the scaling limit of the Hermitian structures performed in~\cite{Bazhanov:2020}.

Here we have only considered the square
lattice models. One could also define
solvable inhomogeneous six-vertex models on
arbitrary planar lattices formed by intersecting straight
lines~\cite{Baxter:1978xr}.
Some of these are related to
the quantizations of circle
patterns~\cite{Bazhanov:2007mh}
arising in the context of a discrete counterpart of the
Riemann mapping theorem~\cite{Bobenko:2004}.
It would be interesting to explore the
scaling limit of these constructions and their connection to CFT.
Finally note that more general lattice systems with an alternating set of
inhomogeneities,
which still belong to the integrability class of the six-vertex model,
were used in the construction of
discrete integrable versions of the sine-Gordon model~\cite{Bobenko:1992rr,Faddeev:1994xy,Bazhanov:1995zg}.

\subsection*{Acknowledgments}
The authors thank R.J.~Baxter for providing
details of the Bethe ansatz for the six-vertex model
on the $45^\circ$-rotated square lattice~\cite{Baxter:1970unpb} and N.Yu.~Reshetikhin for important comments.
VB ack\-now\-ledges the support of the Australian Research Council grant DP180101040.
The~research of~GK is funded by the Deutsche Forschungsgemeinschaft (DFG, German Research
Foundation) under Germany's Excellence Strategy~-- EXC 2121 ``Quantum Universe''~-- 390833306.
The~research of SL is supported by the
Rutgers New High Energy Theory Center.

\pdfbookmark[1]{References}{ref}
\LastPageEnding

\end{document}